\documentclass[twocolumn]{aastex62}

\usepackage[normalem]{ulem}
\useunder{\uline}{\ul}{}
\usepackage{hyperref}
\usepackage{lineno}

\usepackage{color}
\usepackage{amsmath,amssymb}

\hypersetup{linkcolor=cyan,citecolor=magenta,filecolor=yellow,urlcolor=blue}

\shorttitle{GRB 171205A}
\shortauthors{}

\begin{document}

\title{GRB 171205A: Hypernova and Newborn Neutron Star}


\author{Yu~Wang}
\affiliation{ICRA, Dip. di Fisica, Universit\`a  di Roma ``La Sapienza'', Piazzale Aldo Moro 5, I-00185 Roma, Italy}
\affiliation{ICRANet, Piazza della Repubblica 10, I-65122 Pescara, Italy} 
\affiliation{INAF -- Osservatorio Astronomico d'Abruzzo,Via M. Maggini snc, I-64100, Teramo, Italy}

\author{L.~M.~Becerra}
\affiliation{Escuela de F\'isica, Universidad Industrial de Santander, A.A.678, Bucaramanga, 680002, Colombia }
\affiliation{ICRANet, Piazza della Repubblica 10, I-65122 Pescara, Italy}

\author{C.~L.~Fryer}
\affiliation{Center for Theoretical Astrophysics, Los Alamos National Laboratory, Los Alamos, NM, 87545, USA}
\affiliation{Computer, Computational, and Statistical Sciences Division, Los Alamos National Laboratory, Los Alamos, NM, 87545, USA}
\affiliation{The University of Arizona, Tucson, AZ 85721, USA}
\affiliation{Department of Physics and Astronomy, The University of New Mexico, Albuquerque, NM 87131, USA}
\affiliation{The George Washington University, Washington, DC 20052, USA}

\author{J.~A.~Rueda}
\affiliation{ICRA, Dip. di Fisica, Universit\`a  di Roma ``La Sapienza'', Piazzale Aldo Moro 5, I-00185 Roma, Italy}
\affiliation{ICRANet, Piazza della Repubblica 10, I-65122 Pescara, Italy}
\affiliation{ICRANet-Ferrara, Dip. di Fisica e Scienze della Terra, Universit\`a degli Studi di Ferrara, Via Saragat 1, I-44122 Ferrara, Italy}
\affiliation{Dip. di Fisica e Scienze della Terra, Universit\`a degli Studi di Ferrara, Via Saragat 1, I-44122 Ferrara, Italy}
\affiliation{INAF, Istituto di Astrofisica e Planetologia Spaziali, Via Fosso del Cavaliere 100, 00133 Rome, Italy}

\author{R.~Ruffini}
\affiliation{ICRA, Dip. di Fisica, Universit\`a  di Roma ``La Sapienza'', Piazzale Aldo Moro 5, I-00185 Roma, Italy}
\affiliation{ICRANet, Piazza della Repubblica 10, I-65122 Pescara, Italy}
\affiliation{INAF,Viale del Parco Mellini 84, 00136 Rome, Italy}

\email{yu.wang@inaf.it, laura.marcela.becerra@gmail.com, \\fryer@lanl.gov, jorge.rueda@icra.it, ruffini@icra.it}

\begin{abstract}
GRB 171205A is a low-luminosity, long-duration gamma-ray burst (GRB) associated with SN 2017iuk, a broad-line type Ic supernova (SN). It is consistent with being formed in the core-collapse of a widely separated binary, which we have called the binary-driven hypernova (BdHN) of type III. The core-collapse of the CO star forms a newborn NS ($\nu$NS) and the SN explosion. Fallback accretion transfers mass and angular momentum to the $\nu$NS, here assumed to be born non-rotating. The accretion energy injected into the expanding stellar layers powers the prompt emission. The multiwavelength power-law afterglow is explained by the synchrotron radiation of electrons in the SN ejecta, powered by energy injected by the spinning $\nu$NS. We calculate the amount of mass and angular momentum gained by the $\nu$NS, as well as the $\nu$NS rotational evolution. The $\nu$NS spins up to a period of $47$ ms, then releases its rotational energy powering the synchrotron emission of the afterglow. The paucity of the $\nu$NS spin explains the low-luminosity characteristic and that the optical emission of the SN from the nickel radioactive decay outshines the optical emission from the synchrotron radiation. From the $\nu$NS evolution, we infer that the SN explosion had to occur at most $7.36$ h before the GRB trigger. Therefore, for the first time, the analysis of the GRB data leads to the time of occurrence of the CO core-collapse leading to the SN explosion and the electromagnetic emission of the GRB event.
\end{abstract}

\keywords{gamma-ray bursts: general -- black hole physics -- pulsars}

\section{Introduction} 
\label{sec:1}


{The Burst Alert Telescope of the Neil Gehrels Swift Observatory onboard (Swift-BAT)} triggered and located GRB 171205A at $07:20:43$ UT on December $17$, $2017$. Swift's X-Ray Telescope (XRT) began to observe $144.7$~s after the BAT trigger \citep{2017GCN.22177....1D}. Soon, \citet{2017GCN.22178....1I} found that the burst was located in a nearby galaxy at redshift $z=0.0368$, which was later confirmed by {the X-shooter telescope of the  Very Large Telescope (VLT/X-shooter) } \citep{2017GCN.22180....1I}. About $5$ d after, the associated type Ic supernova (SN) started to emerge and was detected by the $10.4$-m {Gran Telescopio Canarias (GTC)} \citep{2017GCN.22204....1D} and SMARTS $1.3$-m telescope \citep{2017GCN.22192....1C}.

\begin{figure*}[t]
    \centering
        \includegraphics[width=0.8\hsize,clip]{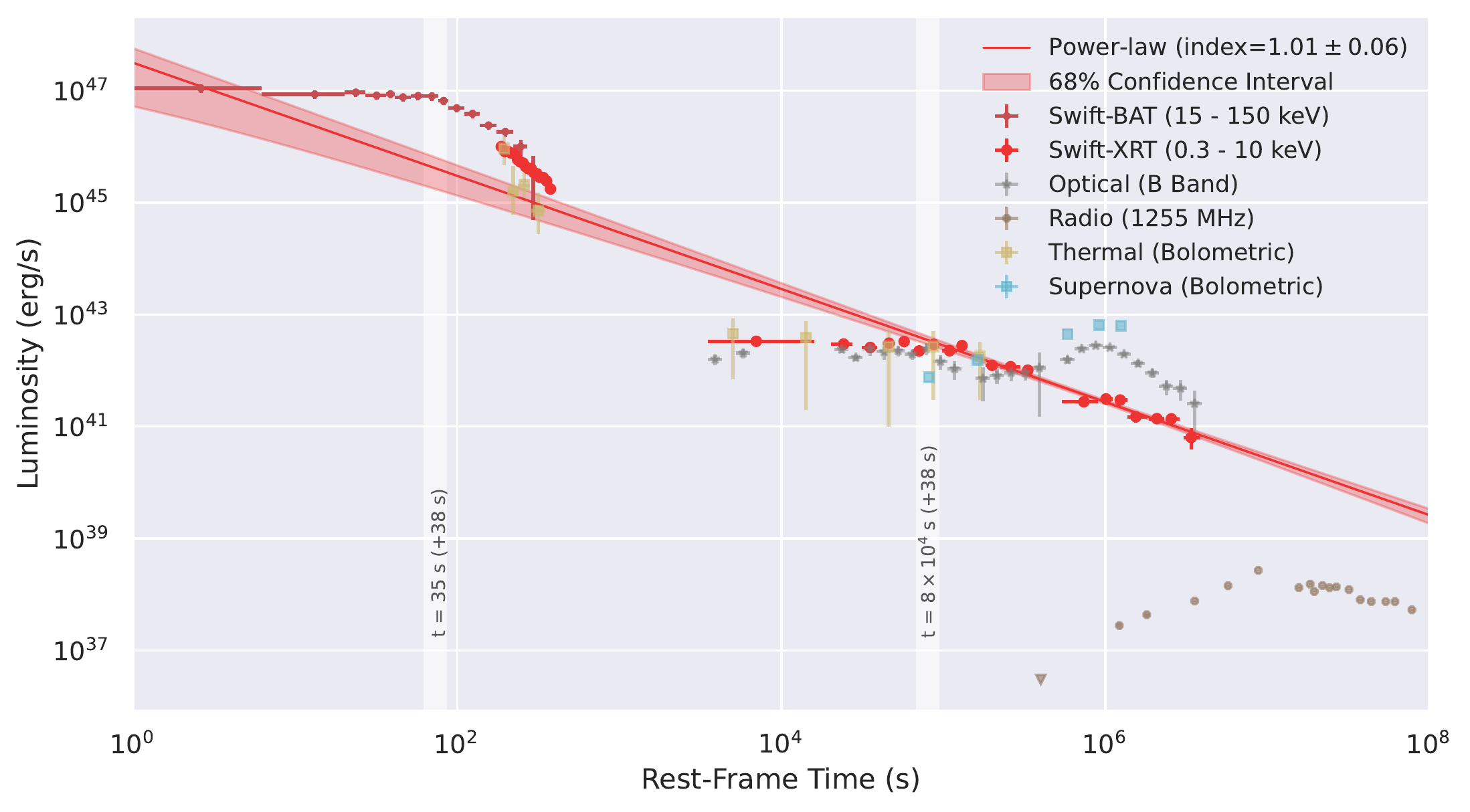}
    \caption{Luminosity light-curve of Swift-BAT (deep red), Swift-XRT (red), optical B band from \citet{2018A&A...619A..66D} (grey), and radio $1255$~MHz from \citet{2021ApJ...907...60M} (brown), the triangles represent the upper limit. We also plot the thermal luminosity (yellow). The Swift-XRT data at time $>8 \times 10^4$~s is fitted by a power-law of index {$1.01\pm0.06$} and extrapolated to the earlier and the later time (red solid line, the red shadow represents the $68\%$ confidence interval). Here $T_0 = 0$~s is the starting time of the burst, corresponds to $38$~s before the BAT trigger time.}
    \label{fig:LCGRB171205A}
\end{figure*}

This source has gained much observational attention since it was the third nearest GRB at the time of its discovery. \citet{2018A&A...619A..66D} performed the multi-wavelength analysis of GRB 171205A using the data from the Swift and Konus-Wind satellites, covering from the optical to the sub-MeV energies. Their cutoff power-law fit gives the peak energy at $\sim 100$~keV and the isotropic energy in the order of $10^{49}$~erg, which implies this burst is a low luminosity GRB and is an outlier of the Amati relation. \citet{2018ApJ...867..147W} reported the spectroscopic observation of the SN associated with the GRB, SN 2017iuk, and of the host galaxy. These observations showed that SN 2017iuk is a typical type Ic SN that resembles SN 2006aj, and that the host is an early-type, star-forming galaxy of high mass, low star formation rate, and low solar metallicity. In this source, for the first time, it was observed the polarization in the millimeter and radio bands during the afterglow phase, thanks to the intensive combined use of {the Submillimeter Array (SMA), the Atacama Large Millimeter/submillimeter Array (ALMA) and the Very Large Array (VLA)}, which shows a linear polarization $<1\%$ indicative of Faraday depolarization \citep{2019ApJ...884L..58U, 2020ApJ...895...64L}. The observation continued for years, the ASKAP, ATCA and $\rm \mu$GMRT radio observations lasted till $\sim 1000$~d, the radio afterglow decays following a shallow power-law and no jet break was exhibited \citep{2021MNRAS.503.1847L, 2021ApJ...907...60M}. Figure \ref{fig:LCGRB171205A} shows the multiwavelength light-curve of GRB 171205A.

{\subsection{GRB 171205A in the traditional scenario}\label{sec:1a}}

The origin of low-luminosity GRBs is still an open debate, and some interpretations include that these are bursts observed off-axis \citep{2004ApJ...602..886W,2006Natur.442.1014S,2006ApJ...638..930S,2016MNRAS.461.1568K,2019ApJ...871..123F,2020A&A...639L..11I}, shockwave breakout from the progenitor's shell \citep{2006Natur.442.1008C,2016MNRAS.460.1680I,2007MNRAS.375..240L,2008Natur.453..469S,2015MNRAS.448..417B,2019ApJ...871..200F}, and emission from a jet-heated cocoon \citep{2015ApJ...807..172N,2017Sci...358.1559K,2018MNRAS.479..588G}. GRB 171205A, as a low luminosity GRB at a low redshift, provides a testing ground for the theoretical models. \citet{2019Natur.565..324I} found thermal X-ray and optical emissions radiated from material whose velocity evolves from $\sim 0.3~c$ to $0.1~c$ in the first $7$ d, and with a chemical composition that differs from that of SN 2017iuk which has a lower velocity ($<0.1~c$) evidenced by the spectroscopic analysis. They proposed the high-velocity material is a portion of the accelerated cocoon, which becomes transparent at $\sim 7$ d, and then the SN dominates the optical emission. \citet{2022ApJ...925..148S} performed hydrodynamic simulations of a powerful jet penetrating the progenitor star and showed that jet-induced chemical mixing can lead to the observed chemical composition of the high-velocity material. \citet{2021ApJ...907...60M} analyzed GRB 171205A with the shockwave breakout and the canonical off-axis jet models and show that both are inconsistent with the $1000$ d observations. Compared to the observation, the shockwave breakout model predicts a longer duration, a lower peak energy, and requires a higher column density. Moreover, the radius ($\sim 10^{13}$~cm) derived from the thermal component is too large for a typical progenitor. For the off-axis model, the discrepancies arise because the burst does not exhibit expected off-axis properties like a low peak energy, a luminosity increasing in the afterglow, and a frequency-independent break in the light-curve \citep{2018A&A...619A..66D}. There are alternative models, e.g., \citet{2019ApJ...870...38S} modeled the burst as mild-relativistic spherical ejecta interacting with an ambient wind-like medium producing forward and reverse shocks and forming a thin shell. In their model, the prompt gamma-ray and X-ray emissions are produced when the optical depth of the shell reaches transparency, and subsequently, the radio and X-ray emissions are produced in the shock fronts by synchrotron and inverse Compton processes. They claimed this model can fit the prompt luminosity and duration, as well as the late-time X-ray, optical, and radio light-curves.

{\subsection{The BdHN scenario}\label{sec:1b}}

Therefore, a satisfactory explanation of the multiwavelength data and the evolution with time of GRB 171205A remains an open issue. In this work, we analyze this source from the perspective of the binary-driven hypernova (BdHN) model of long GRBs. The progenitor of the GRB in the BdHN model is a binary system composed of a carbon-oxygen (CO) star and a neutron star (NS) companion. Numerical simulations of the sequence of physical processes occurring in a BdHN has been performed in the last decade and have led to a detailed picture and interpretation of the GRB observables \citep[see, e.g.,][]{2012ApJ...758L...7R, 2012A&A...548L...5I, 2014ApJ...793L..36F, 2015PhRvL.115w1102F, 2015ApJ...812..100B, 2016ApJ...833..107B, 2018ApJ...852...53R, 2019ApJ...871...14B}. The core-collapse of the CO star leads to the formation of a newborn NS ($\nu$NS) at its center and ejects the outer layers of the star in a SN explosion. The ejecta accretes onto the NS companion and due to matter fallback there is also accretion onto the $\nu$NS {(see, e.g., \citealp{PhysRevD.106.083002, 2022ApJ...936..190W, PhysRevD.106.083004}, and references therein)}. Both accretion processes are hypercritical (i.e., highly super-Eddington) in view of the activation of a very efficient neutrino emission \citep{2016ApJ...833..107B, 2018ApJ...852..120B}. For orbital periods of a few minutes, the NS companion reaches the critical mass for gravitational collapse, leading to a Kerr black hole (BH). These BdHN have been called of type I (BdHN I). BdHN I explain the energetic GRBs with isotropic energies $\gtrsim 10^{52}$ erg. The accretion processes are observed as precursors of the prompt emission \citep[see, e.g.,][]{2019ApJ...874...39W}. The gravitomagnetic interaction of the newborn Kerr BH with the surrounding magnetic field induces an electric field. For a sufficiently supercritical magnetic field, the electric field becomes also supercritical leading to an electron-positron ($e^+e^-$) pair plasma. The self-acceleration of this plasma to Lorentz factors $\Gamma \sim 100$ and its transparency explain the ultra-relativistic prompt emission (UPE) phase {(see \citealp{2021PhRvD.104f3043M} for the UPE analysis of GRB 190114C, and \citealp{2022EPJC...82..778R} for GRB 180720B)}. The electric field accelerates electrons to ultra-relativistic energies leading to synchrotron radiation that explain the observed GeV emission \citep{2019ApJ...886...82R, 2020EPJC...80..300R, 2021A&A...649A..75M, 2022ApJ...929...56R}. There is an additional synchrotron radiation process by relativistic electrons in the ejecta expanding in the $\nu$NS magnetic field. The $\nu$NS also injects energy into the ejecta. This synchrotron radiation explains the afterglow emission in the X-rays, optical, and radio wavelengths \citep[see, e.g.,][]{2018ApJ...869..101R, 2019ApJ...874...39W, 2020ApJ...893..148R}. Finally, the release of nickel decay (into cobalt) in the SN ejecta powers the bump observed in the optical in the late afterglow. 

For longer orbital periods, of the order of tens of minutes, the NS companion does not reach the critical mass, so it remains as a massive, fast rotating NS. These BdHN have been called of type II (BdHN II). BdHN II explain the less energetic GRBs with isotropic energies $\lesssim 10^{52}$ erg. The physical processes and related observables associated with the presence of the BH are clearly not observed in the BdHN II (e.g., the UPE and the GeV emission). The synchrotron afterglow in the X-rays, optical, and radio wavelengths, instead, is present both in BdHN I and II because it is powered by the $\nu$NS and the SN ejecta (see \citealp{2019ApJ...874...39W,2022ApJ...936..190W} for GRB 180728A and GRB 190829A).

{\subsection{GRB 171205A and the quest for BdHN III}\label{sec:1c}}

When considering BdHN with longer and longer orbital period, possibly of hours, the effects associated with the presence of the binary companion become observationally irrelevant. Therefore, there is no GRB observable that can discriminate the presence or absence of a binary companion. Under the above circumstances, we model GRB 171205A neglecting the observational consequences of a companion NS. We shall call these low-luminous sources with energies $\lesssim 10^{49}$--$10^{50}$ erg, as BdHNe III.

Table \ref{tab:observables} summarizes the sequence of physical phenomena that occur in BdHN I, II, and III and their corresponding observables in the GRB data. Signatures from a binary companion appear only in BdHN I and II, while BdHN III shows only observables associated with the SN and the $\nu$NS.

\begin{table*}
{
    \centering
    {
    \caption{Physical phenomena that occur in BdHN I, II and III, and their associated observables in the GRB data.}
    \begin{tabular}{l|c|c|c|c|c|c}
    \hline
     Physical phenomenon & BdHN &\multicolumn{5}{c}{GRB observable} \\
     \cline{3-7}
          & type & $\nu$NS-rise & UPE & GeV & SXFs & Afterglow\\
          & &  (soft-hard X-rays)       & (MeV) & emission & HXFs & (X/optical/radio)\\
          \cline{1-7}
    Early SN emission$^a$ & I, II, III  & $\bigotimes$ & & & & \\
    \cline{1-1}
    Hypercritical accretion onto $\nu$NS$^b$  & I, II, III & $\bigotimes$ & & & & \\
    \cline{1-1}
    Hypercritical accretion onto NS$^b$  & I, II &$\bigotimes$ & & & & \\
      \cline{1-1}
     BH formation from NS collapse$^c$ & I & & & $\bigotimes$ & & \\ 
     \cline{1-1}
    Transparency of $e^+e^-$ (from vacuum & I & & $\bigotimes$ & & &\\
    polarization) with low baryon load region$^d$ & & & & & & \\
     \cline{1-1}
    Synchrotron radiation \textit{inner engine}: & I  & & & $\bigotimes$ & &\\
    BH + $B$-field+SN ejecta$^e$ & & & & & & \\
     \cline{1-1}
    Transparency of $e^+e^-$ (from vacuum & I & & & & $\bigotimes$ &\\
    polarization) with high baryon load$^f$  & & & & & &\\
     \cline{1-1}
    Synchrotron emission from SN ejecta with & I, II, III & & & & & $\bigotimes$\\
    energy injection from $\nu$NS$^g$ & & & & & & \\
     \cline{1-1}
    Pulsar-like emission from $\nu$NS$^g$ & I, II, III & & & & & $\bigotimes$\\
    \cline{1-7}
    \end{tabular}
    }
    \tablecomments{UPE stands for ultrarelativistic prompt emission, SXFs for soft X-ray flares, and HXFs for hard X-ray flares.}
    \tablerefs{$^a$Aimuratov et al. (Submitted to ApJ), \cite{2019ApJ...874...39W, 2022ApJ...936..190W, PhysRevD.106.083004},$^b$\cite{ 2014ApJ...793L..36F, 2016ApJ...833..107B, PhysRevD.106.083002, PhysRevD.106.083004, 2022ApJ...936..190W}, $^c$\cite{2019ApJ...886...82R, 2021A&A...649A..75M, 2021PhRvD.104f3043M},     $^d$\cite{2001A&A...368..377B, 2021PhRvD.104f3043M, 2022EPJC...82..778R}, $^e$\cite{2019ApJ...886...82R, 2020EPJC...80..300R, 2021A&A...649A..75M, 2022ApJ...929...56R}, $^f$\cite{2018ApJ...852...53R}, $^g$\cite{2018ApJ...869..101R, 2019ApJ...874...39W, 2020ApJ...893..148R}.}
    \label{tab:observables}
    }
\end{table*}

In Sec. \ref{sec:2}, we analyze the Swift observations and fit the time-resolved spectra using the MCMC method, then we generate the light-curves for the prompt emission and afterglow, shown in Figs. \ref{fig:LCGRB171205A} and \ref{fig:spectrumXRTBAT}. The special feature of this burst is the presence of a thermal component in the early afterglow, where the temperature drops from about $90$~eV to $70$~eV in the first $300$~s. In Sec. \ref{sec:3}, we describe the physical process of this burst, we suggest that this low-luminosity burst originates from a strong SN (or a hypernova). The fallback accretion after the SN collapse heats up the SN ejecta, accelerating its outermost layer to mild-relativistic and the heated ejecta emits thermal radiation. This process is similar to the cocoon model, but the opening angle for the energy release of the fallback accretion is much larger than the traditional jet. This large opening angle is consistent with the absence of the jet break signal in the afterglow. In the meanwhile, the fallback accretion spins up the central NS, which in turn injects energy to power the afterglow by losing its rotational energy. In Sec. \ref{sec:4}, we establish the analytical solutions for the spin-up of the $\nu$NS due to the mass and angular momentum transfer during the accretion. We derive an analytical solution for the time required for the spin-up process using an accurate Padè approximant in the expression of the angular velocity as a function of time (see Figs. \ref{fig:omvst} and \ref{fig:pade}). The spin period of the NS required by the theory can be obtained from the observation by assuming that the energy of the X-ray afterglow is mainly contributed by the rotational energy of the {$\nu$NS}. From the observation of GRB 171205A, we derive that the {$\nu$NS} is possibly accelerated to a spin period of $47$~ms, and $0.026~M_\odot$ are accreted by the $\nu$NS via fallback. We show that this process takes $7.36$ h for a $\nu$NS born with zero spin. In Sec. \ref{sec:5}, we present the model of the afterglow in the X-rays, optical, and radio wavelengths as originating from synchrotron radiation in the expanding SN ejecta with the energy injection from the central $47$~ms spinning $\nu$NS pulsar. Section \ref{sec:6} shows the results of the fit of the X, optical and radio light-curves with the above model (see Fig. \ref{fig:fit171205A}). The conclusions are given in Sec. \ref{sec:7}.


\section{Spectrum and light-curve}
\label{sec:2}

Swift-BAT and Swift-XRT data are retrieved from UKSSDC \footnote{\url{http://www.Swift.ac.uk}}, the data reduction are performed by Heasoft 6.29 \footnote{\url{http://heasarc.gsfc.nasa.gov/lheasoft/}}, then the exported spectra are fitted by the Multi-Mission Maximum Likelihood framework (3ML) \citep{2015arXiv150708343V}. In order to produce the luminosity light-curve, the BAT data are binned following the thresholds that the signal to noise ratio (SNR) is at least 6 and the maximal bin size is at most $50$~s. Then each binned spectrum is fitted by a cutoff power-law (CPL) function and is integrated from  $15$~keV to $150$~keV according to the BAT bandwidth to obtain the flux. After having the fitting parameters, the fluxes and by adopting the FRW cosmology\footnote{The Friedman-Lema\^itre-Robertson-Walker metric is used for computing the luminosity distance, Hubble constant $H_0=67.4\pm0.5$~km/s/Mpc, and matter density $\Omega_M = 0.315\pm0.007$ \citep{2018arXiv180706209P}.}, the k-corrected luminosity light-curve is obtained \citep{2001AJ....121.2879B}. We generate the light-curve of XRT in the energy range $0.3$--$10$~keV following a similar procedure, the corresponding binning thresholds change to at least $200$ counts and $10$ s duration for the windows timing (WT) mode, as well as at least $100$ counts and $100$ s duration per bin for the photon counting (PC) mode. All the XRT spectra are fitted by a power-law function\footnote{To have more data points for the light-curve, our binning is more concerned with sufficiently short time resolution than with exact spectra. Therefore, the power-law model is used uniformly to fit the spectra, rather than the more accurate power-law plus blackbody model for which the data of each small bin cannot constrain all parameters. This introduces an error of less than $5\%$, which is in an acceptable level.} with the photoelectric absorption models of our Galaxy and the host galaxy. The generated Swift luminosity light-curves are presented in Fig. \ref{fig:LCGRB171205A}. We notice that this burst is seen since $\sim 38$~s before the BAT trigger, hence we set $T_0$ as $38$~s before the BAT trigger time. The XRT light-curve later than $8\times10^4$~s is fitted by a power-law function using \textit{lmfit} \citep{matt_newville_2021_5570790}, a python package for non-linear optimization and curve fitting. {\textit{lmfit} implements the Levenberg-Marquardt method for optimization and is extended by \textit{numdifftool}\footnote{\url{https://numdifftools.readthedocs.io}} to estimate the covariance matrix then to calculate parameter uncertainties.} We obtain a power-law index $1.01\pm0.06$ {with the 1-$\sigma$ uncertainty ($68\%$ confidence level). We show the power-law fit in Fig. \ref{fig:LCGRB171205A} with the 1-$\sigma$ uncertainty region.} The extrapolation of the power-law function coincides with the initial prompt luminosity.

The $T_{90}$ of the BAT observation lasts $189.19$~s, its time-integrated can be described by a cutoff power-law model with power-law index {$\alpha = 1.10\pm0.35$}, while the peak energy cannot be precisely constrained $E_p=148.55\pm 121.97$~keV. These parameters are consistent with \citet{2018A&A...619A..66D}, which jointly fitted BAT and Konus-\textit{Wind} data. They obtained $\alpha=0.85^{+0.54}_{-0.41}$ and $E_p = 122^{+111}_{-32}$~keV, where the uncertainty of peak energy has been tightened because Konus-\textit{Wind} covers higher energies than BAT. The integrated flux gives $(1.56\pm 0.31) \times 10^{-8}$~erg~cm$^{-2}$~s$^{-1}$ in the observed $15$--$150$~keV bandwidth, and extrapolated to $(2.63\pm 0.54) \times 10^{-8}$~erg~cm$^{-2}$~s$^{-1}$ in $1$--$10^4$~keV, which corresponds to the isotropic energy  $E_{\rm iso} = (1.71\pm 0.35) \times 10^{49}$~erg.

The presence of a thermal component in the afterglow of GRB 171205A has been reported in several articles \citep{2017GCN.22191....1C,2018A&A...619A..66D,2019Natur.565..324I}. Our time-resolved analysis also confirms that the additional thermal component significantly improves the fit to the low-energy band of the XRT ($<1$ keV) till $324$~s with a fitting blackbody temperature that drops from $\sim 90$~eV to $\sim 70$~eV, with an uncertainty of $\sim 10$~eV. Afterward, the thermal spectrum gradually fades out of the XRT band ($0.3$--$10$ keV) as the temperature decreases. The WT data of XRT is unable to constrain the temperature at a times later than $\sim 4000$~s, while the optical telescopes start to capture the thermal component that cools to the optical band \citep{2019Natur.565..324I}. 

\begin{figure*}[t]
    \centering
        \includegraphics[width=0.8\hsize,clip]{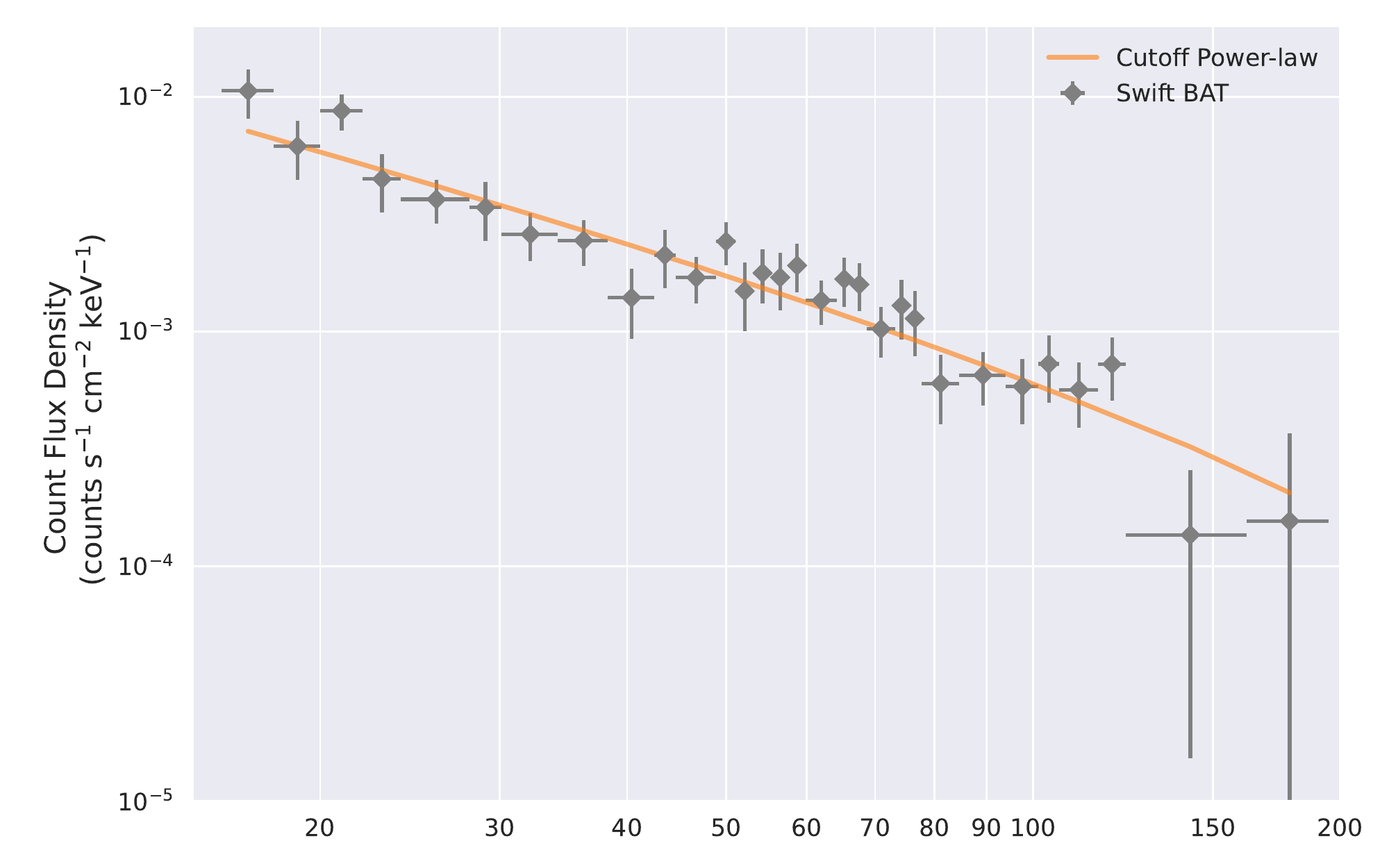}
        \includegraphics[width=0.8\hsize,clip]{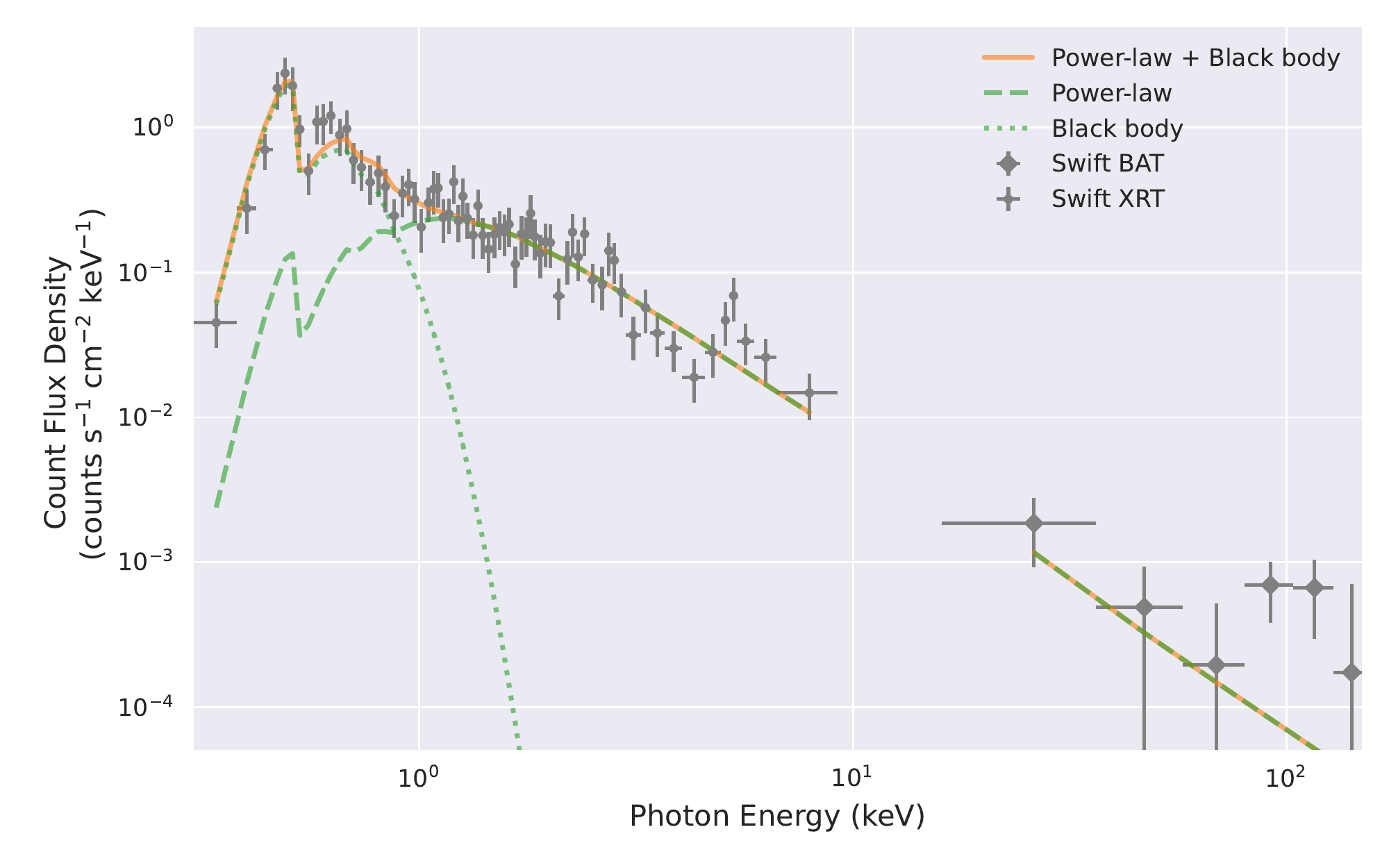}
    \caption{\textbf{Top}: Spectrum of $T_{90}$ observed by BAT,  fitted by a cutoff power-law model with photon index {$\alpha = 1.10\pm0.35$} and peak energy $E_p=148.55\pm121.97$~keV. \textbf{Bottom}: Jointly spectral fitting of BAT and XRT from $151$~s to $162$~s after the BAT trigger with a composite spectrum of power-law of index $\alpha = 2.00\pm0.17$ plus a blackbody of temperature $kT = 77.48\pm7.46$ eV.}
    \label{fig:spectrumXRTBAT}
\end{figure*}

There is a common time window for BAT and XRT observing the source, from $\sim 151$~s when XRT had slewed to the GRB position, till $\sim 162$~s, the end of the $T_{90}$ of BAT. The BAT data at the end of the prompt emission is adequate to constrain the cutoff energy, hence the model of a power-law of index { $\alpha = -2.00\pm0.17$} plus a blackbody component of $kT = 77.53\pm 8.28$~eV is implemented to fit the entire data, as shown in Fig. \ref{fig:spectrumXRTBAT}. 

The optical and radio light-curves in Fig. \ref{fig:LCGRB171205A} are reproduced from \citet{2018A&A...619A..66D} and \citet{2021ApJ...907...60M}, respectively. The optical luminosity is unusually bright compared to the X-rays. \citet{2019Natur.565..324I} found that the evolution of the optical spectrum before and after 7 days is dominated by two blackbodies with different evolution laws. The {1000-day} radio light-curve shows a shallow decay without any jet break signature. We refer to \citet{2018A&A...619A..66D, 2019Natur.565..324I,2021ApJ...907...60M} for the detailed analyses and discussion of the optical and radio data, including the SN optical observation.

\section{Physical Picture}
\label{sec:3}

At a given moment, a type Ic SN occurs from the core-collapse of the CO star, forming at the same time a $\nu$NS at its center. The fallback accretion spins up the $\nu$NS (see Sec. \ref{sec:4}), while releasing the accretion energy. From \citet{2019ApJ...871...14B}, the initial accretion rate is up to a few of $10^{-3}~M_\odot$~s$^{-1}$ and lasts tens of seconds, then it drops following a power-law depending on the SN density profile. Therefore, in the initial phase of tens of seconds, the total energy generated from the accretion and to be injected into the stellar shells reaches $\sim 10^{52}$~erg, which is comparable to the kinetic energy of SN ejecta inferred from the optical emissions at a later time. Different from the traditional jetted model of GRBs, this amount of energy is emitted in a large opening angle of probably tens of degrees, it propagates in a portion of shells and accelerates the outermost shell to the mild-relativistic velocity. The hydrodynamics can be referred to the simulation in \citet{2018ApJ...852...53R}, where has been simulated the propagation of GRB injected energy in the expanding stellar shells. The Lorentz factor of the shockwave is lower than $5$ when it breaks out the outermost shell at $\sim 10^{12}$~cm.  The acceleration of the accretion-powered blastwave is similar to that proposed for the shock-accelerated GRB model~\citep{1974ApJ...187..333C}.  In this scenario, a supernova blastwave accelerates as it propagates down the steep density gradient at the edge of a massive star~\citep{1974ApJ...187..333C,2001ApJ...551..946T}.  Although these models can produce highly-relativistic ejecta in idealized conditions, the bulk of the material reaches only mildly relativistic velocities.  Our model mirrors this evolution, differing only from this picture because the blastwave is propagating through an exploding CO star and is not spherical.  Our asphericity has many of the features of the cocoon produced in jet models \citep[see e.g.][]{2001ApJ...556L..37M,2002MNRAS.337.1349R,2004ApJ...608..365Z,2017ApJ...834...28N,2022MNRAS.517..582E,2022arXiv220804875S}, that the jet pushes the stellar shells sideways to form a hot cocoon, a part of the cocoon emerges from the shells and expands outward with a mild-relativistic velocity. Hence, both our picture and the cocoon picture involve some heated high-velocity material originated from the stellar shells expanding and emitting a thermal spectrum. The evolution of such this blackbody spectrum has been indeed observed by Swift-XRT and several optical telescopes, and a mass of $1.1\times 10^{-3} M_\odot$ moving above $10^5$~km~s$^{-1}$ has been inferred; see Fig. \ref{fig:spectrumXRTBAT} and \citet{2019Natur.565..324I}. The difference is that in our picture, we expect a wider opening angle than in a jet, as we consider this low-luminosity GRB originates from a strong SN or hypernova in which the central compact object is the $\nu$NS. From the observations, there is no signature of any jet break in the afterglow till $\sim 1000$~days \citep{2021MNRAS.503.1847L, 2021ApJ...907...60M}, hence preferring a large opening angle description. 

At this stage, our system has three energy sources; the accretion, the spinning $\nu$NS, and the high-velocity material. For the prompt emission, this low-luminosity GRB deviates from the Amati relation \citep{2002A&A...390...81A}; its peak energy ($E_p=148.55$~keV, see Fig. \ref{fig:spectrumXRTBAT}) is about one order of magnitude higher than the typical value of a weak GRB with isotropic energy $\sim 10^{49}$~erg \citep{2018A&A...619A..66D}. The deviation indicates this burst could be an extreme case or is formed by a different mechanism. \citet{2019Natur.565..324I} suggests that the jet deposits the majority energy in the creation of the cocoon and only a small fraction of energy emitted in gamma-rays. In our framework, accretion dominates the energy release once the SN explodes, and the majority of energy is injected into the stellar shells, converting to the internal and kinetic energy of the SN ejecta, and producing the fast moving material. The low isotropic energy of the prompt emission can be either produced by the tail of accretion or by the fast moving material \citep{2018MNRAS.478.4553D}. For the X-ray afterglow, it can be accounted for, at early times, by the synchrotron emission converted from the kinetic energy of the fast moving material, and at times after the plateau, by the release of rotational energy of the $\nu$NS that has been spun up to periods of the order of milliseconds. We have performed the numerical fitting of the spectrum and light-curve using this scenario for several GRBs \citep[see, e.g.,][]{2018ApJ...869..101R,2019ApJ...874...39W,2020ApJ...893..148R}. This is also supported by that the ending time of the plateau coincides with the transparency of the fast moving material at $\sim 10^5$~s. For the optical afterglow, we share the same opinion with \citet{2019Natur.565..324I}, that the fast expanding mass dominates the optical emission before $4$ days, then the dominance is overtaken by photons diffused out from the massive SN ejecta heated by the nickel radioactive decay.

The above picture contains many different physical processes, most of which have been discussed in detail and simulated, as the references mentioned in the text. However, after the birth of $\nu$NS, the fallback accretion, the mass change and the spin-up process have been rarely discussed in GRB studies. Hence, we will focus on modelling the properties of the newborn NS in the next section.

\section{Spin-up and fallback accretion onto the $\nu$NS}
\label{sec:4}

We turn now to estimate the spin-up and the amount of mass that the $\nu$NS has accreted to gain enough rotational energy to power the X-ray afterglow emission, as specified in the BdHN model \citep[see, e.g.,][for the analysis of 380 BdHNe]{2021MNRAS.504.5301R}.

Assuming the X-ray luminosity as a good proxy of the bolometric luminosity of the afterglow, we can estimate the change in the $\nu$NS rotational energy from a time $t_1$ to a time $t_2 > t_1$ from the energy balance equation, i.e.
\begin{equation}\label{eq:Erotvst}
    \int_{t_1}^{t_2} \dot{E}_{\rm rot}\,dt = E_{\rm rot}(t_2) - E_{\rm rot} (t_1) \approx -\int_{t_1}^{t_2} L_X dt.
\end{equation}

After infinite time, the $\nu$NS will have lost all its rotational energy, therefore when $t_2 \to \infty$, we have $E_{{\rm rot},\infty} (t_2) \to 0$. So, assuming the time $t_1$ to be a generic time $t$, and the power-law luminosity 
\begin{equation}\label{eq:Lx}
    L_X = A_X t^{-\alpha_X}, 
\end{equation}
we obtain from Eq. (\ref{eq:Erotvst}) that the $\nu$NS angular velocity evolves as
\begin{equation}\label{eq:Omvst}
    \Omega (t) \approx \sqrt{\frac{2 A_X\,t^{1-\alpha_X}}{(\alpha_X-1) I}},
\end{equation}
where $I$ is the stellar moment of inertia which we have assumed constant with time, and can be estimated, for instance, using the EOS-independent approximate expression \citep{2019JPhG...46c4001W}
\begin{equation}\label{eq:ILS}
I \approx \left( \frac{G}{c^2} \right)^2 M^3 \sum_{i=1}^4\frac{b_i}{(M/M_\odot)^i},
\end{equation}
where $b_1 = 1.0334$, $b_2 = 30.7271$, $b_3 = -12.8839$, and $b_4 = 2.8841$.

In the case of GRB 171205A, the X-ray luminosity is fitted by a power-law at times $t>t_{\rm pl}\approx 8\times 10^4$ s, with {$A_X = (3.165\pm 0.238)\times 10^{47}$ erg s$^{-1}$, and $\alpha_X = 1.022 \pm 0.055$}. Using these values, we estimate from Eq. (\ref{eq:Omvst}) that the rotation period of the $\nu$NS at $t=t_{\rm pl}$ is {$P (t_{\rm pl}) \approx 51.01$ ms}. If we assume that the $\nu$NS is spinning down from the $\nu$NS-rise, i.e., from $t = t_{\nu \rm NS} \approx 35$ s, but the emission from it is partially absorbed by the high-velocity material which is opaque before $\sim 10^5$~s, then by extrapolating from $t=t_{\rm pl}$ backward in time to $t = t_{\nu \rm NS}$, we infer that at the $\nu$NS-rise time, the $\nu$NS rotation period was {$P_{\nu\rm NS} \equiv P (t_{\nu \rm NS}) \approx 46.85$ ms, i.e., $\Omega (t_{\nu \rm NS}) = 134.11$ rad s$^{-1}$}.

We now estimate the mass accreted by the $\nu$NS before the $\nu$NS-rise, so to spin it up to the above rotation rate. The accretion rate onto the $\nu$NS, set by the amount of mass from the inner layers of the expanding matter that fallback onto the $\nu$NS and their infalling speed, proceeds at hypercritical rates \citep[see, e.g.,][]{1996ApJ...460..801F}. The accretion process makes the $\nu$NS to increase its mass-energy and rotation rate from the transfer of baryonic mass and angular momentum. The evolution of the $\nu$NS gravitational mass and angular momentum can be calculated from \citep{2019ApJ...871...14B}
\begin{align}
      \dot{M}&=\left( \frac{\partial M}{\partial M_b} \right)_{J} \, \dot{M}_b + \left( \frac{\partial M} {\partial J}\right)_{M_b}\, \dot{J},\label{eq:Mdot}\\
      \dot{J}&= \tau_{\rm acc},\label{eq:Jdot}
\end{align}
where $J = I \Omega$ is the angular momentum, $M$ is the gravitational mass, $M_b$ the baryonic mass, $\dot{M}_b$ is the baryonic mass accretion rate, and $\tau_{\rm acc}$ is the accretion torque. 

Equation (\ref{eq:Mdot}) must be complemented with the expressions of the two partial derivatives. These relations can be calculated from the fitting formula of the NS binding energy obtained in \citet{2015PhRvD..92b3007C}
\begin{equation}\label{eq:MbMns}
  \mu_b - \mu = \frac{13}{200}\mu^2\left(1-\frac{1}{130}j^{1.7} \right),
\end{equation}
where $j\equiv cJ/(GM_\odot^2)$ is the dimensionless angular momentum and $\mu = M/M_\odot$. From it, we redily obtain
\begin{align}
    \left(\frac{\partial \mu}{\partial \mu_b} \right)_{j} &= \frac{1}{1+\frac{13}{100}\mu\left(1-\frac{1}{130}j^{1.7}\right)},\\ 
    \left( \frac{\partial \mu} {\partial j}\right)_{\mu_b} &= \frac{\frac{1.7}{2000}\mu^2 j^{0.7}}{1+\frac{13}{100}\mu\left(1-\frac{1}{130}j^{1.7}\right)}.
\end{align}

The numerical simulations of BdHNe performed in \citet{2019ApJ...871...14B} show that the material accreted by the $\nu$NS circularizes around it in a sort of Keplerian disk structure before being accreted. Therefore, we assume that the accreted matter exerts onto the $\nu$NS the torque
\begin{equation}\label{eq:chi}
  \tau_{\rm acc} = \chi\,l\,\dot{M}_b,
\end{equation}
where $l$ the specific (i.e. per unit mass) angular momentum of the innermost stable circular orbit around the $\nu$NS, and $\chi \leq 1$ is an efficiency parameter of angular momentum transfer. For the angular momentum of the last stable circular orbit, we use the approximate EOS-independent results presented in \citet{2017PhRvD..96b4046C}
\begin{equation}
  l = 2\sqrt{3}\frac{G M}{c}\left[1 \mp 0.107\left( \frac{j}{M/M_\odot} \right)^{0.85}\right].
  \label{eq:lISO}
\end{equation}

We can obtain an approximate, analytic solution to Eq. (\ref{eq:Jdot}). For this task, we use the following analytic formula that fits the numerical results of the fallback accretion rate calculated in \citet{2019ApJ...871...14B, PhysRevD.106.083002}
\begin{equation}\label{eq:Mbdot}
    \dot{M_b} \approx \dot{M}_0 \left(1+\bar{t} \right)^{-p},
\end{equation}
where $\dot{M}_0 = 7.2 \times 10^{-4} M_\odot$ s$^{-1}$, $t_{\rm acc} = 12$ s, $p = 1.3$, and we have introduced the notation $\bar{t} = t/t_{\rm acc}$. 

\begin{figure}
    \centering
    \includegraphics[width=\hsize,clip]{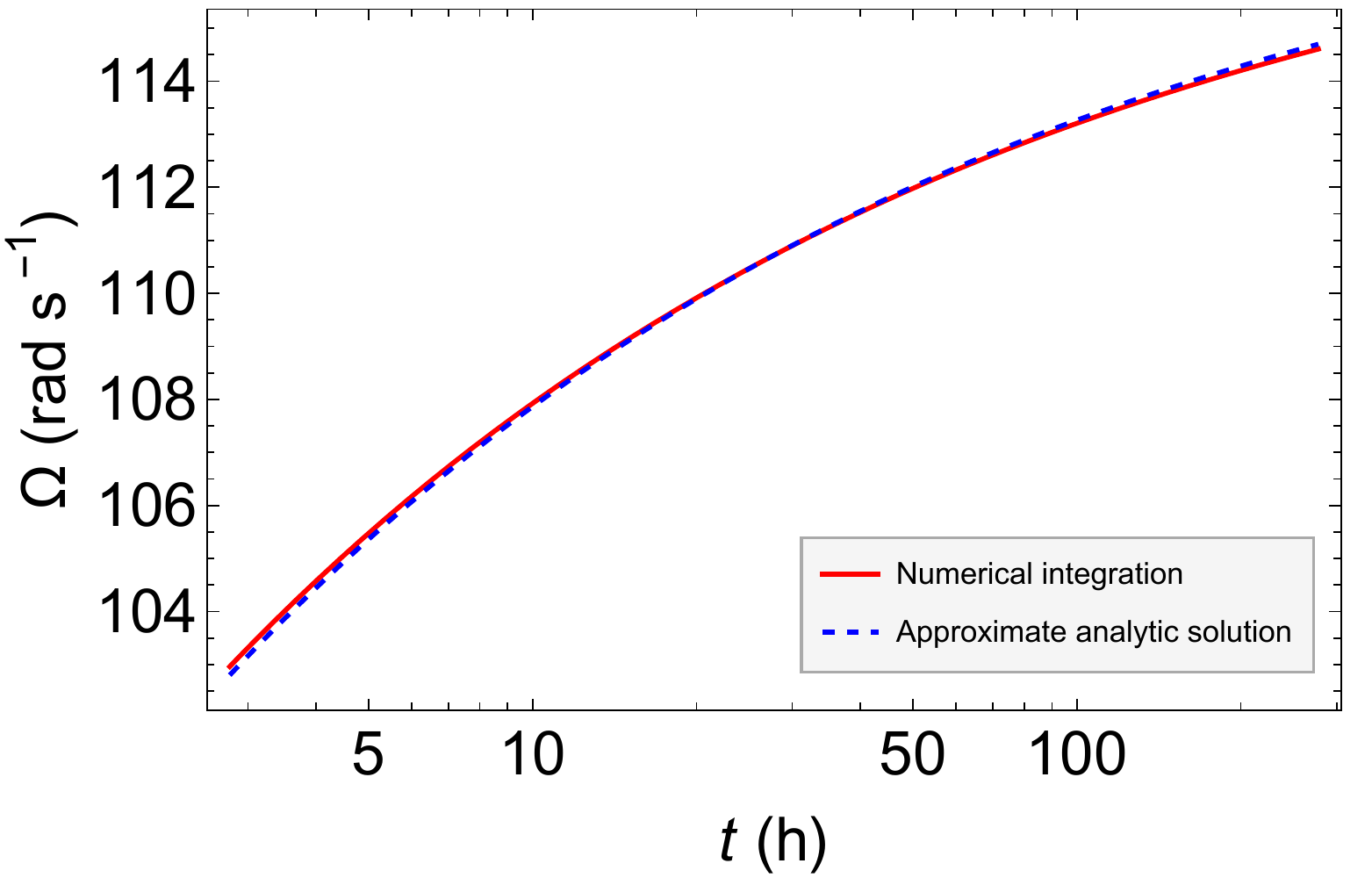}
    \caption{Comparison of the approximate solution of the Eqs. (\ref{eq:Mdot}) and (\ref{eq:Jdot}) given by Eq. (\ref{eq:DeltaOmega}), with the results from the full numerical integration, in the case $\mu(t_0)=1.4$, $\Omega(t_0)=0$, and $\chi=0.15$.}
    \label{fig:omvst}
\end{figure}

For the involved rotation rates ($j \sim 0.01$), the contribution of the rotation terms in Eqs. (\ref{eq:MbMns}) and (\ref{eq:lISO}) is negligible, so we can retain only the first term in those equations. With this assumption, and integrating Eq. (\ref{eq:Mbdot}), we have
\begin{align}
    \mu_b &= \mu_b(t_0) + \frac{\dot{M}_0 t_{\rm acc}}{p-1}\left[ 1 - \left( 1 + \bar{t} \right)^{1-p} \right],\label{eq:mub}\\
    \mu &\approx \frac{100}{13}\left(\sqrt{1+\frac{13}{50}\mu_b} - 1 \right),\label{eq:mu}\\
    l &\approx 2 \sqrt{3}\frac{G M_\odot}{c} \mu, \label{eq:japp}
\end{align}
where $\mu_b(t_0) \approx \mu_0 + (13/200) \mu_0^2$, being $\mu_0 = M(t_0)/M_\odot$ the initial $\nu$NS gravitational mass, and we have inverted Eq. (\ref{eq:MbMns}) to write the gravitational mass in terms of the baryonic mass. Equations (\ref{eq:mub}) and (\ref{eq:mu}) implies that in the limit $t\to \infty$ the baryonic mass and the gravitational mass approaches a maximum value
\begin{align}
    \mu_{b,\rm max} &= \mu_b(t_0) + \frac{\dot{M}_0 t_{\rm acc}}{p-1} = \mu_b(t_0) + 0.0288,\label{eq:mubmax}\\
    \mu_{\rm max} &= \frac{100}{13}\left(\sqrt{1+\frac{13}{50}\mu_{b,\rm max}} - 1 \right).\label{eq:mumax}
\end{align}
We now approximate the angular momentum derivative as $\dot{J} \approx I \dot{\Omega} \approx I_{\rm max} \dot{\Omega}$, where $I_{\rm max} = I (\mu_{\rm max})$, so that Eq. (\ref{eq:Jdot}) becomes
\begin{equation}\label{eq:Omdot}
    \dot{\Omega} \approx \beta \mu(t) (1+\bar{t})^{-p},\quad \beta = \frac{2 \sqrt{3} G M_\odot^2 \chi \dot{\mu}_0}{c I_{\rm max}},
\end{equation}
whose solution can be written as
\begin{equation}\label{eq:Omegasol}
    \Omega(t) = \Omega(t_0) + \beta\int_{t_0}^t \mu(t) (1+\bar{t})^{-p} dt.
\end{equation}
Making the change of variable $x = (1+\bar{t})^{1-p}$, the integration of Eq. (\ref{eq:Omegasol}) is straightforward leading to
\begin{align}\label{eq:DeltaOmega}
    &\Delta \Omega = \Omega(t) -\Omega(t_0) \nonumber\\
    &=\omega \left\{ x + \frac{2}{3} k \left[ \left(1+\frac{13\mu_b}{50}  \right)^{3/2} - \alpha^{3/2}\right] - 1 \right\},
\end{align}
where we have defined
\begin{align}\label{eq:constants}
    \omega &= \frac{100}{13}\frac{\beta\, t_{\rm acc}}{p-1},\quad \Delta \mu_b = \frac{\dot{M}_0 t_{\rm acc}}{p-1} = 0.0288,\\
    k &= \frac{50}{13}\frac{1}{\Delta \mu_b} = 133.547,\quad \alpha = 1+\frac{13}{50}\mu_{b,0}
\end{align}
and we have set the initial time $t_0 = 0$ since the fallback accretion begins soon after the SN explosion \citep[see, e.g.,][]{2019ApJ...871...14B}. Figure \ref{fig:omvst} compares the approximate analytic solution (\ref{eq:DeltaOmega}) with the solution from the full numerical integration of Eqs. (\ref{eq:Mdot}) and (\ref{eq:Jdot}), in the case of $\mu(t_0)=1.4$, $\Omega(t_0)=0$, and $\chi=0.15$.

Equation (\ref{eq:DeltaOmega}) tells us that in the limit $t \to \infty$ ($x \to 0$), the $\nu$NS reaches asymptotically a maximum angular velocity gain
\begin{equation}\label{eq:DeltaOmegamax}
    \Delta \Omega_{\rm max} = \omega \left\{\frac{2}{3} k \left[ \left(1+\frac{13\mu_{b,\rm max}}{50}  \right)^{3/2} - \alpha^{3/2}\right] - 1 \right\},
\end{equation}
which as expected is larger for larger values of the angular momentum transfer efficiency parameter, $\chi$. Since we assume that after the $\nu$NS-rise the $\nu$NS is spinning down, we seek for solutions with a spinning up phase that ends with an angular velocity approaching the value that we have inferred at the $\nu$NS-rise, i.e.
\begin{equation}\label{eq:constraint}
    \Omega_{\rm max} \approx \Omega(t_{\rm \nu NS}),
\end{equation}
where $\Omega_{\rm max} = \Delta \Omega_{\rm max} + \Omega(t_0)$. We have used the approximate symbol in Eq. (\ref{eq:constraint}) because by definition the value $\Omega_{\rm max}$ is reached only asymptotically. For practical purposes, we seek for solutions in which $\Omega(t_{\rm \nu NS})=0.9\,\Omega_{\rm max}$. Therefore, given values of $M$ and $\Omega(t_{\rm \nu NS})$, the above constraint leads to a specific value of $\chi$ that leads to the self-consistent spin-up phase. For instance, for a $\nu$NS mass $M=1.4 M_\odot$ and
{
\begin{equation}\label{eq:OmnuNS}
    \Omega_{\nu\rm NS}\equiv \Omega(t_{\rm \nu NS}) = \sqrt{\frac{2 A_X\,t_{\nu\rm NS}^{1-\alpha_X}}{(\alpha_X-1) I}} \approx 134.11\,\,\text{rad s}^{-1},
\end{equation}
we obtain $\chi = 0.182$}.

We can also obtain a simple analytic estimate of the mass accreted by assuming that during the spin up phase, the accretion rate, the gravitational mass, and the moment of inertia are constant and have their maximum values. Under this assumption, Eqs. (\ref{eq:Jdot}) and (\ref{eq:chi}) lead to the accreted mass in a time $\Delta t$,
\begin{equation}\label{eq:deltaMb}
    \Delta \mu_b \approx \frac{c I_{\rm max} \Delta \Omega}{2 \sqrt{3} \chi G M_\odot^2 \mu_{\rm max}}.
\end{equation}
For the above parameters, Eq. (\ref{eq:deltaMb}) gives $\Delta \mu_b \approx 0.02570$. This is very close to the value obtained from the full numerical integration, $\Delta \mu_b = 0.02592$, which represents an error of only $0.85\%$. The accuracy of Eq. (\ref{eq:deltaMb}) resides in the fact that the fallback accretion rate decreases as a power-law, see Eq. (\ref{eq:Mbdot}), hence most of the baryonic mass is accreted in the first minutes of the evolution. This explains why the above value of the accreted mass is close to the maximum accreted mass given by Eq. (\ref{eq:mubmax}), i.e., $\Delta \mu_{b, \rm max} = 0.0288$. 

\begin{figure}
    \centering
    \includegraphics[width=\hsize,clip]{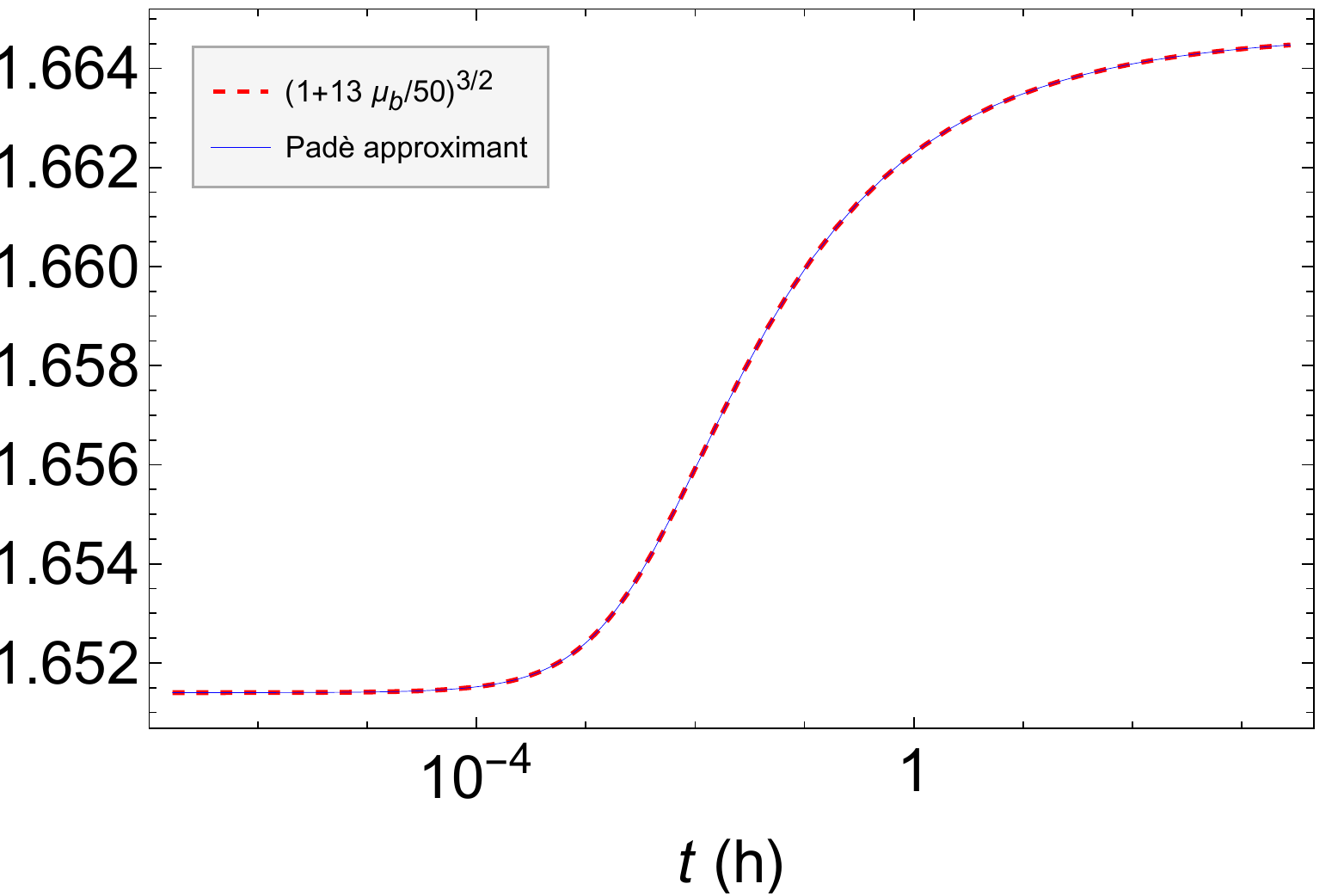}
    \caption{Comparison of the Padè approximant given by Eq. (\ref{eq:Pade}) with the result of the full numerical integration.}
    \label{fig:pade}
\end{figure}

We turn to obtain an analytic expression of the time interval $\Delta t$ elapsed since the beginning of the fallback accretion, up to the instant when the $\nu$NS reaches a given angular velocity, or a given angular velocity gain, $\Delta \Omega$. In principle, we can obtain it by inverting Eq. (\ref{eq:DeltaOmega}). However, the equation is highly non-linear, so to obtain a relatively simple expression for it we use an accurate Padè approximant for the quantity involving the baryonic mass, i.e.

\begin{align}
    \left(1+\frac{13\mu_b}{50}  \right)^{3/2} &= b^{3/2}(\tilde{\alpha} + X)^{3/2}\approx b^{3/2}({\cal F} + \tilde{\alpha}^{3/2}),\nonumber \\
    {\cal F} &= \frac{6\tilde{\alpha}^{3/2} X}{4\tilde{\alpha} - X},\label{eq:Pade}
\end{align}

where $b=1/k=(13/50)\Delta \mu_{b,\rm max}$, $\tilde{\alpha} = \alpha/b$, and we have introduced the variable $X=1-x$. For the same example of Fig. \ref{fig:omvst}, we show in Fig. \ref{fig:pade} the excellence performance of the Padè approximant (\ref{eq:Pade}), which approximate the expression with a tiny error of only $10^{-9}$. 

Using the approximant (\ref{eq:Pade}), Eq. (\ref{eq:DeltaOmega}) becomes a second-order polynomial in the variable $X$ whose solution is straightforward, leading to the time interval:
\begin{equation}\label{eq:Deltat}
    \Delta t = t_{\rm acc} \left[\left(1-X\right)^{\frac{1}{1-p}}-1\right],
\end{equation}
where

\begin{align}
    X &= \frac{B + \sqrt{B^2 + 4 C}}{2}\label{eq:X}\\
    B &= 4 \tilde{\alpha} - 4 \tilde{\alpha}^{3/2} \sqrt{b}-\Delta \Omega/\omega, \label{eq:B}\\
    C &= 4 \tilde{\alpha}\Delta \Omega/\omega. \label{eq:C}
\end{align}

The relevance of the above time interval is that it allows to compute the time elapsed to reach the angular velocity at the $\nu$NS-rise, $\Omega (t_{\nu\rm NS})$. Since it is close to the maximum value reachable by the fallback accretion, that time interval gives an estimate of the time elapsed since the SN explosion {(SN-rise)}, $t_{\rm SN}$. For the present example, we obtain 
{
\begin{equation}\label{eq:tSN}
    t_{\rm SN} = \Delta t (\Delta \Omega) = \Delta t(\Omega_{\nu \rm NS}) \approx 7.36\,\,\text{h},
\end{equation}
where we have used $\Delta \Omega = \Omega_{\nu \rm NS} - \Omega(t_0) = 134.11$ rad s$^{-1}$, as given by Eq. (\ref{eq:OmnuNS})}. The full numerical integration leads to $7.20$ h, which implies that Eq. (\ref{eq:Deltat}) estimates the time interval with an error of only $2.2\%$.

\section{Synchrotron and pulsar emission}\label{sec:5}

We turn now to the specific modeling of the multiwavelength afterglow of GRB 171205A. In the present scenario, the non-thermal component of the afterglow originates from the synchrotron radiation in the SN ejecta. The SN ejecta gets energy injected from the $\nu$NS fallback accretion and the multipolar emissions. Numerical calculations of this model applied to the description of the afterglow of specific GRBs can be found in \citet{2018ApJ...869..101R, 2019ApJ...874...39W, 2020ApJ...893..148R}. An analytic treatment of the model has been presented in \citet{2022arXiv220200316R}, and \citet{2022arXiv220705619W} have applied it to model the afterglow of GRB 180720B. {Our afterglow model relies more on the continuous energy injections than the traditional forward shockwave model that relies on the kinetic energy of jet. And unlike the traditional model that only considers the injection of dipole emission as an additional energy source to explain the short duration plateau (for e.g. internal plateau) \citep{{1998A&A...333L..87D,1998PhRvL..81.4301D,2001ApJ...552L..35Z,2011MNRAS.413.2031M,2012PhRvD..86j4035L,2017ApJ...849..119C,2018ApJS..236...26L,2020ApJ...896...42Z}}, our modelling process takes into account the fallback accretion, the dipole and quadrupole radiation, such continuous energy injections produce the long-lasting afterglow.} We here follow the latter to estimate for GRB 171205A the emission generated by the synchrotron mechanism in the X-rays, in the optical, and in the radio, and the $\nu$NS pulsar emission.

\subsection{Synchrotron emission by the expanding ejecta}\label{sec:5a}

The distribution of radiating electrons per unit energy, $N(E,t)$, is obtained from the solution of the kinetic equation \citep{1962SvA.....6..317K}
\begin{equation}\label{eq:kinetic}
    \frac{\partial N(E, t)}{\partial t}=-\frac{\partial}{\partial E}\left[\dot{E}\,N(E,t)\right] + Q(E,t),
\end{equation}
where $Q(E,t)$ is the number of injected electrons into the ejecta per unit time $t$, per unit energy $E$, and $\dot E$ is the electron energy loss rate. 

Following \citet{2022arXiv220200316R, 2022arXiv220705619W}, we adopt the solution to Eq. (\ref{eq:kinetic}) {for a self-similar uniform expansion}
\begin{align}\label{eq:N3}
&N(E,t)\approx \begin{cases}
    \frac{q_0}{\beta B_{*,0}^2 (\gamma-1)}\hat{t}^{2} E^{-(\gamma+1)}, & t < t_q\\
   \frac{q_0 (t_q/t_*)^{k}}{\beta B_{*,0}^2 (\gamma-1)}\hat{t}^{2-k} E^{-(\gamma+1)}, &   t_q < t < t_b,
\end{cases}
\end{align}
where $E_b < E < E_{\rm max}$, being
\begin{equation}\label{eq:Eb}
    E_b = \frac{\hat{t}}{{\cal M} t_*},\quad
    t_b = {\cal M} t_*^2 E_{\rm max}.
\end{equation}
The model parameters are defined as follows. The ejecta expands self-similarly with the radiating layer being $r=R_* = R_{*,0}\,\hat{t}$, $\hat{t} \equiv t/t_*$, $t_* = R_*/v_* = R_{*,0}/v_{*,0}$, $v_* = R_*(t)/t = v_{*,0}$, $B_*(t) = B_{*,0} R_{*,0}/r = B_{*,0}\hat{t}^{-1}$ is the magnetic field strength at $r=R_*$, ${\cal M}\equiv \beta B^2_{*,0}/2$, $\beta = 2e^4/(3 m_e^4 c^7)$. We assume the injection power-law distribution $Q(E,t)=Q_0(t)E^{-\gamma}$ \citep{1962SvA.....6..317K, 1973ApJ...186..249P, 1979rpa..book.....R, 2011hea..book.....L}, where $\gamma$ and $E_{\rm max}$ are parameters to be determined from the observational data, and $Q_0(t)$ can be related to the power released by the $\nu$NS and injected into the ejecta from 
$L_{\rm inj}(t)=L_0 (1+t/t_q)^{-k} = \int_{0}^{E_{\rm max}} E\,Q(E,t) dE$, so $Q_0(t) = q_0\left(1+t/t_q\right)^{-k}$, where $q_0 \equiv  (2-\gamma)L_0/E_{\rm max}^{2-\gamma}$.

The bolometric synchrotron radiation power of a single electron is given by \citep[see, e.g.,][]{2011hea..book.....L}
\begin{equation}\label{eq:Psyn}
    P_{\rm syn}(E,t) = \beta B_*^2(t) E^2 \approx \frac{\beta}{\alpha} B_* \nu,
\end{equation}
where in the last equality we have used the fact that most of the radiation is emitted at frequencies near the so-called critical frequency, $\nu_{\rm crit} = \alpha B_* E^2$, where $\alpha = 3 e/(4\pi m_e^3 c^5)$. {By setting $N(E,t) = \eta\,\hat{t}^l E^{-p}$, so with the constants $\eta$ , $l$ and $p$ obtained by comparing this expression with Eq. (\ref{eq:N3}),} the synchrotron luminosity radiated at frequencies from $\nu_1$ to $\nu_2 > \nu_1$ can be written as
\begin{align}\label{eq:Lnu}
    L_{\rm syn}(\nu_1,\nu_2; t) &= \int_{\nu_1}^{\nu_2} J_{\rm syn}(\nu,t)d\nu\approx \nu J_{\rm syn}(\nu,t),\nonumber \\
    &
    \approx \frac{\beta}{2} \alpha^{\frac{p-3}{2}} \eta B_{*,0}^{\frac{p+1}{2}}\hat{t}^{\frac{2 l- n(p+1)}{2}}\nu^{\frac{3-p}{2}}.
\end{align}
where $\nu_1=\nu$, $\nu_2=\nu+\Delta\nu$, being $\Delta\nu$ the bandwidth. Here, $J_{\rm syn}$ is the spectral density which is given by $J_{\rm syn}(\nu,t)d\nu\approx P_{\rm syn}(\nu,t) N(E,t)dE$ \citep[see, e.g.,][]{2011hea..book.....L}. In Eq. (\ref{eq:Lnu}), we have made the approximation $\Delta\nu/\nu\ll 1$ because of the power-law character of the spectral density. Despite the synchrotron radiation of a single electron is beamed along the velocity of the particle, we here consider an isotropic distribution of a large number of electrons with an isotropic distribution of pitch angles, hence leading to an isotropic total synchrotron luminosity. 

\subsection{Newborn NS evolution and pulsar emission}\label{sec:5b}

The $\nu$NS is subjected to the angular momentum loss driven by the magnetic field braking. In the point dipole+quadrupole magnetic field model presented in \citet{2015MNRAS.450..714P}, the total magnetic torque is given by
\begin{align}
    \tau_{\rm mag} &= \tau_{\rm dip} + \tau_{\rm quad},\label{eq:taumag}\\
    \tau_{\rm dip} &= -\frac{2}{3} \frac{B_{\rm dip}^2 R^6 \Omega^3}{c^3} \sin^2\alpha,\\
    \tau_{\rm quad} &= -\frac{32}{135} \frac{B_{\rm quad}^2 R^8 \Omega^5}{c^5} \sin^2\theta_1 (\cos^2\theta_2+10\sin^2\theta_2),
\end{align}
where $\alpha$ is the inclination angle of the magnetic dipole moment with respect to the rotation axis, and the angles $\theta_1$ and $\theta_2$ specify the geometry of the quadrupole field. The strength of the magnetic dipole field is $B_{\rm dip}$. The dipole pure axisymmetric mode ($m = 0$) is set by $\alpha = 0$, and the pure $m=1$ mode by $\alpha = \pi/2$. The strength of the quadrupole magnetic field is $B_{\rm quad}$. The quadrupole $m=0$ mode is set by $\theta_1 = 0$, the $m=1$ mode by $\theta_1 = \pi/2$ and $\theta_2=0$, while the $m=2$ mode is set by $\theta_1 = \theta_2 = \pi/2$. For the fit of the data, we shall adopt the $m=1$ mode for the dipole while the quadrupole can range between the $m=1$ and $m=2$ modes. {The existence of multipolar magnetic fields in the newly born NS is supported by some theories and observations \citep{2013Natur.500..312T,2013MNRAS.434.1658M,2016MNRAS.456.4145R,2019LRCA....5....3P}}. Therefore, we can write the total magnetic torque (\ref{eq:taumag}) as
\begin{equation}\label{eq:taumagfinal}
    \tau_{\rm mag} = -\frac{2}{3} \frac{B_{\rm dip}^2 R^6 \Omega^3}{c^3}\left( 1 + \xi^2 \frac{16}{45} \frac{R^2 \Omega^2}{c^2} \right),
\end{equation}
where $\xi$ is the quadrupole to dipole magnetic field strength ratio defined by
\begin{equation}\label{eq:eta}
    \xi \equiv \sqrt{\cos^2\theta_2+10\sin^2\theta_2} \frac{B_{\rm quad}}{B_{\rm dip}},
\end{equation}
and the spindown luminosity as
\begin{equation}\label{eq:Lsd}
    L_{\rm sd} = \Omega\,|\tau_{\rm mag}| = \frac{2}{3} \frac{B_{\rm dip}^2 R^6 \Omega^4}{c^3}\left( 1 + \xi^2 \frac{16}{45} \frac{R^2 \Omega^2}{c^2} \right).
\end{equation}
The evolution of the $\nu$NS is obtained from the energy conservation equation
\begin{equation}\label{eq:Erot}
	-(\dot{W}+\dot{T}) = L_{\rm tot} = L_{\rm inj} + L_{\rm sd},
\end{equation}
where $W$ and $T$ are, respectively, the $\nu$NS gravitational and rotational energy. 

\section{Results}\label{sec:6}

\begin{figure*}
\centering
\includegraphics[width=0.8\hsize,clip]{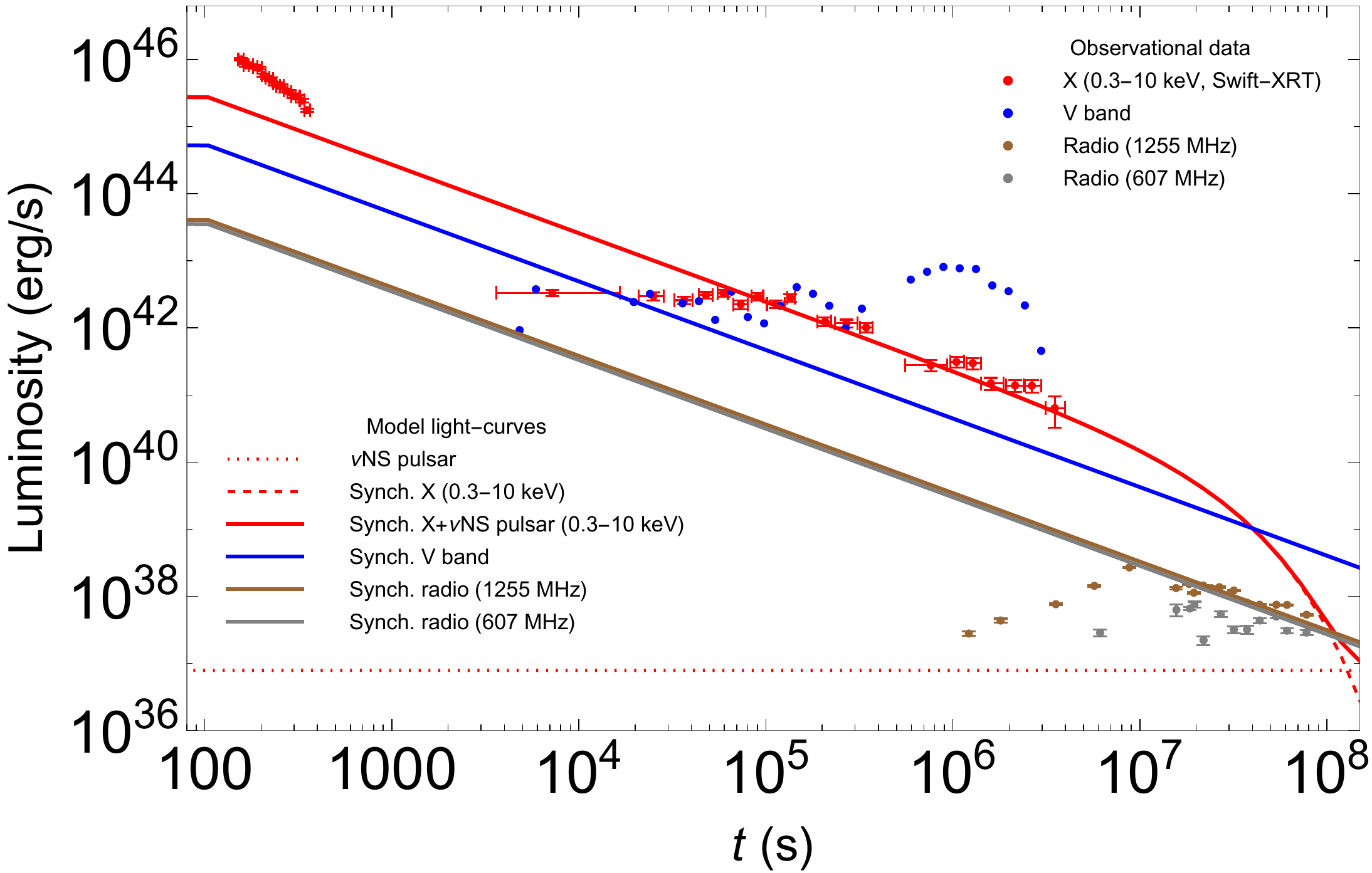}
\caption{Luminosity of GRB 171205A in the X-rays ($0.3$--$10$ keV), optical (V band), and radio ($607$ MHZ and $1.255$ GHz) energy bands compared with the luminosity predicted by the theoretical model. The raising part of the radio luminosity in the time interval $10^6$--$10^7$ s is due to synchrotron self-absorption \citep[see][for details]{2021ApJ...907...60M}, here unmodeled. The X-ray data is retrieved from the Swift-XRT repository and analyzed by this article, the optical and radio data are reproduced from \citet{2018A&A...619A..66D} and \citet{2021ApJ...907...60M}.}
\label{fig:fit171205A}
\end{figure*}

{The emission of GRB 171205A comprises thermal and non-thermal components. In Sec. \ref{sec:1a}, we recalled that \citet{2019Natur.565..324I} explains the thermal component up to $10^5$ s in the X-rays and in the optical due to the cooling of fast moving material. We here address the nature of the non-thermal component once the material is transparent. Therefore, the present model of synchrotron radiation described in Sec. \ref{sec:5}, aims to explain the data that shows a decreasing power-law luminosity in the different energy bands with similar power-law index. 

Table \ref{tab:parameters} {summarizes} the values of the model parameters that fit the afterglow of GRB 171205A in the X-rays, optical, and radio energy bands, as shown in Fig. \ref{fig:fit171205A}, obtained according to the above guidelines and the fitting procedure outlined in Appendix \ref{app:A}.

\begin{table}
{
    \centering
    \begin{tabular}{l|r}
    Parameter & Value \\
    \hline
    $t_{*}$ ($10^{4}$ s) & $2.650 \pm 110.276$ \\
    $B_{*,0}$ ($10^{5}$ G) & $3.774 \pm 157.021$\\
    $\gamma$ &  $1.606 \pm 0.231$\\
    $E_{\rm max}$ ($10^4 \ m_e c^2$) & $3.738$\\
    $k$  & $1.219 \pm 0.170$\\
    $L_0$ ($10^{47}$ erg s$^{-1}$)& $1.011\pm 0.801$\\
    $t_q$ (s) & $100.00$\\
    $B_{\rm dip}$ ($10^{12}$ G) & $1.000$ \\
    $P_{\nu\rm NS}$ (ms) & $46.852 \pm 64.910$\\
       \hline
    \end{tabular}
    \caption{Value of the synchrotron model parameters that fit the multiwavelength observational data of GRB 1701205A as shown in Fig. \ref{fig:fit171205A}.}
    \label{tab:parameters}
    }
\end{table}

In the X-rays, the model describes the decreasing power-law behavior at times $>10^5$ s, and in the radio at times $>10^7 s$. We do not model the rising part of the radio emission in the time interval $10^6$--$10^7$ s which is due to synchrotron self-absorption \citep[see][for details]{2021ApJ...907...60M}}.

The first relevant feature to notice is that the afterglow luminosity fades with time with an approximate power-law $t^{-1}$. This power-law is shallower than in GRBs of higher luminosity in which $t^{-1.3}$ (see, e.g., GRB 130427A or GRB 190114C in \citealp{2018ApJ...869..101R, 2020ApJ...893..148R}). The pulsar emission from magnetic braking predicts a luminosity with a sharper power-law, in a pure magnetic dipole the luminosity falls as $t^{-2}$, and for a pure magnetic quadrupole as $t^{-3/2}$ (see equations of Sec. \ref{sec:5b} and \citealp{2018ApJ...869..101R, 2020ApJ...893..148R}). Therefore, models based on pulsar emission from magnetic braking alone (even including higher order multipole fields) are unable to fit the afterglow luminosity of GRB 171205A. This is a first indication of the necessity of an additional mechanism, in this case the synchrotron radiation. The second relevant feature is that the afterglow in the X-rays and in the radio bands show {similar} power-law index (see the red, gray and brown data points), as expected from the synchrotron model.

The optical data shows, instead, a flat behavior followed by the bump that characterizes the peak of the SN emission powered by the decay of nickel in the ejecta \citep{1996snih.book.....A,2019Natur.565..324I}. {Both the synchrotron radiation and the SN radioactive decay contribute to the optical emission, but in GRB 171205A the latter dominates over the former. This explains the deviation of the optical luminosity from the typical power-law behavior of the synchrotron radiation. This feature is consistent with the BdHN III nature of the source. In fact, BdHN III} are low-luminous sources {in which} the $\nu$NS is not a very-fast rotator, so it injects less energy into the ejecta in comparison to BdHNe I (e.g., GRB 130427A, 180720B or 190114C; see \citealp{2018ApJ...869..101R, 2020ApJ...893..148R}) and BdHNe II (e.g., GRB 190829A; see \citealp{2022ApJ...936..190W}). Therefore, the synchrotron emission is not very luminous and the emergent optical SN outshines the optical synchrotron luminosity. Interestingly, this latter feature of the emergent optical SN emission is also fulfilled in the most general situation of BdHN I and BdHN II (Aimuratov et al., to be submitted). SN 2017iuk is similar to the SNe associated with high-luminous GRBs, indicating that the pre-SN progenitor (i.e. the CO star and a NS companion) leading to the $\nu$NS in its core-collapse event, is similar for all long GRBs irrespective of their energetics (Aimuratov et al., to be submitted).

In the X-rays, the synchrotron luminosity fades off after a few $10^6$ s, when $h \nu_{\rm crit}$ falls below keV. At later times, the power-law behavior continues in the optical and in the radio bands. The pulsar emission is characterized by a plateau followed by a power-law decay (at times longer than the characteristic spindown timescale). For a plateau luminosity comparable (but smaller) to the synchrotron power-law luminosity, the sum of the two contributions can lead to a luminosity with a less sharp power-law behavior than the pure synchrotron. The afterglow of GRB 171205A does not show any sign of change of the power-law of the synchrotron emission (see Fig. \ref{fig:fit171205A}), so we can not constrain the magnetic field strength and structure. In Fig. \ref{fig:fit171205A}, we have adopted {$\approx 47$ ms} as initial rotation period of the $\nu$NS and a pure dipole field ($\xi=0$) of {$B_{\rm dip} = 10^{12}$ G} to guide the eye of the reader. For magnetic fields $\gtrsim 5\times 10^{13}$ G, the plateau luminosity of the pulsar emission contributes appreciably to the total X-ray luminosity affecting the goodness of the fit. Therefore, we can assume the above estimate as an upper limit to the dipole magnetic field. For the present synchrotron model parameters, X-ray data after times of a few $10^6$ s could help to constrain the presence of the pulsar emission. A sanity check of the model is that the energy injected into the ejecta is $\sim 10^{49}$ erg, of the same order of the rotational energy of the $\nu$NS, for a moment of inertia of a few $10^{45}$ g cm$^2$.

\section{Conclusions}
\label{sec:7}

In this article, we have interpreted GRB 171205A within the BdHN model of long GRBs. {In particular, because of the low energy release of only a few $10^{49}$ erg, we have classified GRB 171205A as a BdHN III, systems with long} orbital periods, perhaps of the order of hours, in which the NS companion does not play any role in the cataclysmic event. Most of these binaries are also expected to be disrupted by the SN explosion \citep{2015PhRvL.115w1102F, 2016ApJ...832..136R, 2018ApJ...859...30R}. Under these circumstances, the GRB event is explained by the sole activity of the $\nu$NS and its interaction with the SN ejecta.

{We have here shown} that GRB 171205A is a low-luminous GRB consistent with it being produced in the core-collapse of a single CO star that forms the $\nu$NS and the type Ic SN. There are several new results related to the sequence of physical phenomena occurring in this system and the related GRB observables: 
\begin{enumerate}
    \item 
    The fallback accretion is initially of a few of $10^{-3}M_\odot$~s$^{-1}$ and lasts tens of seconds \citep{2019ApJ...871...14B, PhysRevD.106.083002}. The accretion energy is $\sim 10^{52}$~erg, comparable to the kinetic energy of the SN ejecta. This energy is injected into the ejecta, propagates, and accelerates the outermost shell to the observed mild-relativistic velocity. The hydrodynamics is similar to the case of the expanding SN ejecta with the GRB energy injection presented in \citet{2018ApJ...852...53R}. The Lorentz factor of the shockwave is $\lesssim 5$ when it gets transparency at $\sim 10^{12}$~cm, and emit a thermal spectrum. This scenario explains the prompt emission of GRB 171205A. This is also similar to the cocoon scenario advanced for this source in \citet{2019Natur.565..324I}. Both pictures predict the heating of stellar shells (in our case by the physical process of the fallback accretion originating from the SN explosion and in the other by postulation a unspecified jet) that get boosted to high-velocity and emit a thermal spectrum. The associated blackbody emission has been indeed observed in GRB 171205A, and it has been inferred that $\approx 10^{-3} M_\odot$ of material expand at velocities above $10^5$~km~s$^{-1}$ (see \citealp{2019Natur.565..324I} and Fig. \ref{fig:spectrumXRTBAT}). The main difference between the two models is that in our picture there is not jet. This solutions seems favoured since the associated jet break expected in the afterglow of jetted GRB models is not observed in the data up to the last observations at $\sim 1000$~days \citep{2021MNRAS.503.1847L, 2021ApJ...907...60M}.
    \item 
     Regarding the afterglow emission, we have first inferred from an energy conservation argument, that the $\nu$NS should have started to lose its rotational energy at $t=35$ s after the GRB trigger, i.e., from what we call the $\nu$NS-rise, with a rotation period of {$47$} ms.
     \item
     We have shown that the afterglow of GRB 171205A can not be explained by the sole pulsar emission of the $\nu$NS by magnetic braking, even including higher multipole fields (e.g., quadrupole).
     \item 
     The multiwavelength afterglow is explained by synchrotron radiation emitted by electrons in the expanding SN, which is further powered by energy injected by the $\nu$NS. We have calculated the synchrotron luminosity in the X-rays, the optical and radio wavelengths with an analytic treatment of the above physical situation. We have shown that the X-rays and the radio luminosities follow the expectation from the synchrotron model. The raising part of the radio luminosity in the time interval $10^6$--$10^7$ s is due to synchrotron self-absorption \citep[see][for details]{2021ApJ...907...60M}. The observed optical luminosity shows a flat behavior followed by the bump of the optical SN powered by the energy release in the ejecta of the radioactive decay of nickel into cobalt. We have shown that the synchrotron luminosity in those optical wavelengths lies below the luminosity of the emergent SN optical emission. This implies that the observed optical emission contains the contribution of both the synchrotron radiation and the optical SN.
     \item 
     Another remarkable fact to be highlighted is that SN 2017iuk, a SN associated with the low-luminous GRB 171205A, a BdHN III, shows similar properties (e.g., peak luminosity and peak time) to the SNe associated with high-luminous GRBs (BdHN I and II). This suggests that the pre-SN progenitor (i.e., the CO star) is similar for all long GRBs, irrespective of their energetics (Aimuratov et al., to be submitted).
     \item 
     There is a corollary of the above result. In low-luminous GRBs, i.e., in BdHN III like GRB 171205A, the relatively slow rotation ({$47$ ms} period) of the $\nu$NS implies the less energy injected into the ejecta, hence the low energetics of the associated synchrotron emission. Under these circumstances, the optical emission of the SN powered by the nickel radioactive decay is able to outshine the optical synchrotron luminosity. 
     \item 
     We calculated the evolution of the $\nu$NS mass and angular momentum (assumed to be initially zero) during the fallback accretion process leading to its spinning up to the {$47$} ms rotation period. From this evolution, we have inferred that the SN explosion occurred at most $7.36$ h before the GRB trigger time. This sets an estimate of the time delay between the SN explosion and the electromagnetic emission of the GRB event, assuming a $\nu$NS born with zero spin.
     
\end{enumerate}

\acknowledgements
We thank the Referee for comments and suggestions that helped us in the presentation of the article.
L.M.B. is supported by the Vicerrector\'ia de Investigación y Extensi\'on - Universidad Industrial de Santander Postdoctoral Fellowship Program No. 2022000293.

\appendix 

{\section{Fitting procedure}}\label{app:A}

{
In this appendix, we describe how we set the value of the model parameters from the physical scenario and specific observables, including the attached uncertainties. The parameters to be specified are the index of the electron's energy injection, $\gamma$, the parameters defining the injected power $k$, $L_0$, and $t_q$, the maximum energy of the electrons, $E_{\rm max}$, the self-similar expansion timescale, $t_*$, the magnetic field at the initial time of reference of the expansion, $B_{*.0}$, the $\nu$NS dipole magnetic field strength $B_{\rm dip}$. In Sec. \ref{sec:4}, we have already fixed the $\nu$NS rotation period, $P_{\nu\rm NS}$. Our aim is to estimate the uncertainty in each parameter from the propagation of the 1-$\sigma$ uncertainty of the power-law fit of the X-ray and radio luminosity.
}

Equation (\ref{eq:Lnu}) shows that the signature of the present synchrotron model is the power-law luminosity in the different bands with (ideally) the same power-law index. Therefore, we constrain the synchrotron model parameters using the observational data showing the above property. Figure \ref{fig:fit171205A} shows that the X-ray ($0.3$--$10$ keV) luminosity data behaves as a power-law in the time interval $t \approx (0.87$--$5)\times 10^6$ s, and the radio ($1255$ MHz) data in the time interval $t \approx (2.92$--$8)\times 10^7$ s. The two luminosities are fitted by
\begin{equation}\label{eq:Lfit}
    L_X = A_X\,t^{-\alpha_X},\qquad L_r = A_r\,t^{-\alpha_r}, 
\end{equation}
where $A_X = (3.165 \pm 0.238)\times 10^{47}$ erg s$^{-1}$, $\alpha_X=1.022 \pm 0.055$, $A_r = (4.290 \pm 0.178)\times 10^{42}$ erg s$^{-1}$, and $\alpha_r = 0.616 \pm 0.081$. The uncertainties at 1-$\sigma$ level. To estimate the uncertainties in the value of the model parameters, derived from the above fit, we follow the standard theory of error propagation. For instance, given quantity $f$ that is a function of the independent variables $a_i$, i.e., $f(a_1, a_2, ..., a_n)$, its uncertainty can be estimated as \citep[see, e.g.,][]{ku1966}
\begin{equation}\label{eq:deltaf}
    \delta f = \sum_{i = 1}^n \sqrt{\left| \frac{\partial f}{\partial a_i}\right|^2 (\delta a_i)^2} \approx \sum_{i = 1}^n \left| \frac{\partial f}{\partial a_i}\right| \delta a_i.
\end{equation}
Thus, the uncertainties of the luminosities given by the power-law fits (\ref{eq:Lfit}), at a time $t$, can be estimated by
\begin{equation}\label{eq:dL}
    \delta L_i \approx \left| \frac{\partial L_i}{\partial A_i} \right|\delta A_i + \left| \frac{\partial L_i}{\partial \alpha_i} \right|\delta \alpha_i = 
    \frac{\delta A_{i}}{A_{i,c}} + |-\ln t|\, \delta\alpha_{i}, \qquad i = X,r,
\end{equation}
where $A_{X,c} = 3.165\times 10^{47}$ erg s$^{-1}$, $A_{r,c} = 4.290\times 10^{42}$ erg s$^{-1}$, $\delta A_{X} = 0.238 \times 10^{47}$ erg s$^{-1}$, $\alpha_{X,c} = 1.022$, $\alpha_{r,c} = 0.616$, $\delta A_{r} = 0.178 \times 10^{42}$ erg s$^{-1}$, $\delta \alpha_X = 0.055$, and $\delta \alpha_r = 0.081$.

We turn to the self-similar expansion timescale, $t_*$, for which we must set values for $R_{*,0}$ and $v_{*,0}$. For $v_{*,0}$, we chose a fiducial value according to numerical simulations of the SN explosion \citep[see, e.g.,][]{2016ApJ...833..107B, 2019ApJ...871...14B}, so we set $v_{*,0} =10^8$ cm s$^{-1}$ and there is no propagated uncertainty to calculate for. With the above, we set the expansion timescale and its attached uncertainty

\begin{equation}\label{eq:tstar}
    t_* = \frac{R_{*,0}}{v_{*,0}},\qquad \delta t_* = \frac{\delta R_{*,0}}{v_{*,0}}.
\end{equation}
According to our working assumption of uniform expansion, the inner radius and its uncertainty are
\begin{equation}\label{eq:Rstar}
    R_{*,0} = v_{*,0}\, t_{\rm SN},\qquad \delta R_{*,0} = v_{*,0}\, \delta t_{\rm SN},
\end{equation}
where $t_{\rm SN}$ is the time since SN given by Eqs. (\ref{eq:Deltat}) and (\ref{eq:tSN}), and $\delta t_{\rm SN}$ its uncertainty. For the above expansion velocity $v_{*,0}$, and the time since SN estimated in Sec. \ref{sec:4}, $t_{\rm SN} \approx 2.650 \times 10^4$ s $ \approx 7.36$ h, we have $R_{*,0}\approx 2.65\times 10^{12}$ cm and $t_* = t_{\rm SN} = 2.65 \times 10^4$ s. The uncertainty attached to the time $t_{\rm SN}$ can be estimated as 
\begin{equation}\label{eq:dtSN}
    \delta t_{\rm SN} = \left| \frac{\partial t_{\rm SN}}{\partial \Omega_{\nu \rm NS}} \right|\delta \Omega_{\nu\rm NS},\qquad \frac{\partial t_{\rm SN}}{\partial \Omega_{\nu \rm NS}} = \frac{1}{\omega}\frac{4 \tilde{\alpha} - X}{2 X - B} \frac{t_{\rm acc}}{1-p}\left[(1-X)^{\frac{p}{1-p}}\right],
\end{equation}

where $X$, and $B$ are given by Eqs. (\ref{eq:X})--(\ref{eq:B}), evaluated at the time $t = t_{\nu \rm NS}$ s, so $\Delta \Omega = \Omega_{\nu \rm NS}$. The uncertainty in estimating $\Omega_{\nu\rm NS}$ from Eq. (\ref{eq:OmnuNS}) is given by

\begin{equation}\label{eq:dOmega}
    \delta \Omega_{\nu\rm NS} = \left | \frac{\partial \Omega_{\nu\rm NS}}{\partial A_X}\right| \delta A_X  + \left | \frac{\partial \Omega_{\nu\rm NS}}{\partial \alpha_X}\right| \delta \alpha_X,\quad \frac{\partial \Omega_{\nu\rm NS}}{\partial A_X} = \frac{\Omega_{\nu\rm NS}}{2 A_X}, \quad \frac{\partial \Omega_{\nu\rm NS}}{\partial \alpha_X} = -\frac{\Omega_{\nu\rm NS}}{2}\left(\frac{1}{\alpha_X-1}+\ln t_{\nu\rm NS} \right).
\end{equation}
For the present parameters, i.e, $t_{\nu \rm NS} = 35$ s and $\Omega_{\nu\rm NS} = 134.11$ rad s$^{-1}$, we obtain from Eq. (\ref{eq:dOmega}), $\delta \Omega_{\nu\rm NS} \approx 185.795$ rad s$^{-1}$, so an uncertainty in the rotation period, $\delta P_{\nu\rm NS} = |\partial P_{\nu\rm NS}/\partial \Omega_{\nu\rm NS}|\delta \Omega_{\nu\rm NS} = 2 \pi \delta \Omega_{\nu\rm NS}/\Omega^2_{\nu\rm NS} \approx 64.910$ ms. Using the above in Eq. (\ref{eq:dtSN}), we obtain $\delta t_{\rm SN} \approx 110.276\times 10^4$ s $\approx 306.220$ h. Thus, we get from Eq. (\ref{eq:Rstar}), $\delta R_{*,0} \approx 110.276 \times 10^{12}$ cm, and from Eq. (\ref{eq:tstar}), $\delta t_* = \delta t_{\rm SN}$.

At large distances from the $\nu$NS, we expect the toroidal component of the magnetic field to dominate, which decays with distance as $r^{-1}$ \citep[see, e.g.,][]{1969ApJ...157..869G}. Assuming a toroidal field of the same order as the poloidal field near the $\nu$NS surface, its value at the radius $r=R_{*,0}$ is
\begin{equation}\label{eq:Bstar}
    B_{*,0} \approx B_{\rm dip}\frac{R}{R_{*,0}} = B_{\rm dip}\frac{R}{v_{*,0} t_*},
\end{equation}
where $B_{\rm dip}$ is the strength of the dipole magnetic field and $R$ is the fiducial $\nu$NS radius. We shall see in Sec. \ref{sec:7} that the data does not constraint the dipole field but only set an approximate upper limit of $B_{\rm dip,max} \approx 5 \times 10^{13}$ G. Therefore, we shall adopt a fiducial, conservative magnetic field value $B_{\rm dip} = 10^{12}$ G. By using a fiducial $\nu$NS radius $R = 10^6$ cm, and the value of $R_{*,0}$ given by Eq. (\ref{eq:Rstar}), we obtain $B_{*,0} \approx 3.774 \times 10^5$ G. With the choice (\ref{eq:Bstar}), the attached uncertainty is given by
\begin{equation}\label{eq:dBstar}
    \delta B_{*,0} = B_{*,0}\frac{\delta R_{*,0}}{R_{*,0}}= B_{*,0}\frac{\delta t_*}{t_*} = B_{*,0}\frac{\delta t_{\rm SN}}{t_{\rm SN}} ,
\end{equation}
which leads to $\delta B_{*,0} \approx 157.021 \times 10^5$ G.

{
We now set the index $\gamma$. From Eq. (\ref{eq:Lnu}), we infer that the ratio of the synchrotron luminosity at two frequencies, $\nu_1$ and $\nu_2$, is given by $L_{\rm syn}(\nu_1)/L_{\rm syn}(\nu_2) = (\nu_1/\nu_2)^{\frac{3-p}{2}}$, where $p = \gamma+1$. Therefore, we can constrain the value of the index $\gamma$ using the data in the X-rays and in the radio as
\begin{equation}\label{eq:gamma}
    \gamma = 2 \left[ 1 - \frac{\ln(L_X/L_r)}{\ln(\nu_X/\nu_r)} \right],
\end{equation}
where $L_X$ and $L_r$ are given in Eqs. (\ref{eq:Lfit}). Since the fitted power-laws are not equal, the value of $\gamma$ inferred from Eq. (\ref{eq:gamma}) depends on the time at which we calculate the ratio of the luminosities. Therefore, we adopt for $\gamma$ the value given by the mean $\langle \gamma \rangle = \Delta t^{-1} \int \gamma dt$. We obtained $\langle \gamma \rangle \approx 1.6060$, where we have used $\Delta t \approx 8\times 10^7$ s, $\nu_X = 10$ keV/$h \approx 2.423 \times 10^{18}$ Hz and $\nu_r = 1255$ MHz. From Eq. (\ref{eq:gamma}), the uncertainty in the choice of $\gamma$ can be estimated as

\begin{equation}\label{eq:dgamma}
    \delta \gamma = \left | \frac{\partial \gamma}{\partial L_X}  \right| \delta L_X + \left | \frac{\partial \gamma}{\partial L_r}  \right| \delta L_r =  \frac{2}{\ln(\nu_X/\nu_r)} \left( \frac{\delta L_X}{L_X} + \frac{\delta L_r}{L_r} \right),
\end{equation}
whose mean for the above parameters is $\langle\delta \gamma \rangle = 0.231$.

{
The synchrotron emission peaks around the critical frequency
\begin{equation}\label{eq:nucrit}
    \nu_{\rm crit} = \alpha B_* E^2 = \alpha B_{*,0} E^2\frac{t_*}{t},
\end{equation} 
and then cutoffs exponentially, where $E$ is the electron energy. Since the critical frequency decreases with time, there is a hard-to-soft evolution of the cutoff  and the X-rays data give the strongest constraint. The electrons of maximum energy, $E_{\rm max}$, produce the maximum critical frequency, $\nu_{\rm crit,max}$. By requiring that $\nu_{\rm crit,max} = \nu_X$ at a cutoff time $t_{\rm cut,X} > t_{f,X}$, where $t_{f,X} \approx 3.5 \times 10^6$ s is the time of the last observational X-ray data, we obtain that the maximum electron energy must at least have the value
\begin{equation}\label{eq:Emax}
    E_{\rm max} = \sqrt{\frac{\nu_X\,t_{\rm cut,X}}{\alpha\, t_*\, B_{*,0}}} = \sqrt{\frac{\nu_X\, v_{*,0}\,t_{\rm cut,X}}{\alpha\, B_{\rm dip} R}},
\end{equation}
where in the last equality we have used Eq. (\ref{eq:Bstar}). The cutoff time must allow the power-law luminosity to extend at least up to $t_{f,X}$. Thus, we chose $t_{\rm cut,X}$ such that the exponential cutoff at the time $t=t_{f,X}$ has reduced the power-law X-ray luminosity to one part in a thousand. With this condition, we find $t_{\rm cut,X} \approx 2.418 \times 10^7$ s, so $E_{\rm max} = 3.738 \times 10^4 m_e c^2$. Equation (\ref{eq:Emax}) tells us that $E_{\rm max}$, chosen in this way, depends only on fiducial values that we have set for $v_{*,0}$, $B_{\rm dip}$, $R$ and $t_{\rm cut,X}$, so we can not estimate an attached uncertainty to it.

{
Having set all the above parameters, it remains to set the parameters of the injected power, $L_0$, $k$ and $t_q$. The synchrotron luminosity increases at times $t< t_q$ (see Sec. \ref{sec:5a}) and decreases at times $t>t_q$. The X-rays luminosity shows always a decreasing behavior, so we set $t_q = 100$ s, which roughly corresponds to the initial time of the X-ray data. For the parameters $L_0$ and $k$, we equate the model synchrotron luminosity (\ref{eq:Lnu}) in the case of X-rays with the power-law luminosity (\ref{eq:Lfit}). From this equality, we obtain
\begin{align}
    k &=\frac{2 + 2\alpha_X -\gamma}{2}, \label{eq:k}\\
    L_0 &= 2\frac{A_X}{t_*^{\alpha_X}} \frac{\gamma-1}{2-\gamma}\left( \frac{t_*}{t_q} \right)^k \left( \frac{t_{\rm cut, X}}{t_*} \right)^{\frac{2-\gamma}{2}} = 2A_X \frac{\gamma-1}{2-\gamma}t_q^{-\alpha_X} \left( \frac{t_{\rm cut, X}}{t_q}\right)^{\frac{2-\gamma}{2}} ,\label{eq:L0}
\end{align}
where we have used Eq. (\ref{eq:Emax}). For the present parameters, we obtain $k = 1.219$ and $L_0 = 1.011 \times 10^{47}$ erg s$^{-1}$. Therefore, we can estimate the error of the above quantities as

\begin{align}
    \delta k &= \left | \frac{\partial k}{\partial \alpha_X}  \right| \delta \alpha_X + \left | \frac{\partial k}{\partial \gamma}  \right| \delta \gamma = \delta \alpha_X + \frac{1}{2} \delta \gamma,\\
    \delta L_0 &= \left | \frac{\partial L_0}{\partial A_X}  \right| \delta A_X + \left | \frac{\partial L_0}{\partial \alpha_X}  \right| \delta \alpha_X  + \left | \frac{\partial L_0}{\partial \gamma}  \right| \delta \gamma = L_0 \left[ \frac{\delta A_X}{A_X} + \ln t_q\, \delta\alpha_X + \left| \frac{1}{(\gamma-1)(2-\gamma)} - \frac{1}{2}\ln \left(\frac{t_{\rm cut,X}}{t_q} \right)  \right| \delta\gamma\right],
\end{align}
which read $\delta k = 0.170$ and $\delta L_0 \approx 0.792 L_0 \approx 0.801 \times 10^{47}$ erg s$^{-1}$.

The large uncertainty in the estimate of $t_*$ and $B_{*,0}$ is a consequence of the propagation of the uncertainty of $R_{*,0}$, which arises from the uncertainty in the estimate of the SN time, $t_{\rm SN}$, because it is a sensitive function of $\Omega_{\nu\rm NS}$.

\bibliographystyle{aasjournal}
\bibliography{171205A}

\end{document}